\documentclass{aa}
\usepackage{txfonts}
\usepackage{graphicx}
\usepackage{longtable}

\newcommand\phs{\phantom{$-$}}

\begin{document}

\title{Chemical Abundances for the Outer Halo Cluster Pal~4 from Co-added High-Resolution Spectroscopy}

\author{Andreas Koch\inst{1} 
  \and Patrick C\^ot\'e\inst{2}   
  }
\authorrunning{A. Koch \& P. C\^ot\'e}
\titlerunning{Chemical abundance constraints of Pal 4}
\offprints{A. Koch \email{ak326@astro.le.ac.uk}}

\institute{Department of Physics \& Astronomy, University of Leicester, University Road, Leicester 
LE1 7RH, UK
  \and National Research Council of Canada, Herzberg Institute of Astrophysics, 
  5071 West Saanich Road, Victoria, BC V9E 2E7, Canada 
}
\date{}

\abstract {Chemical element abundances for distant Galactic globular clusters (GCs) hold important clues to the origin of the Milky Way halo and its substructures.} 
                 {We study the chemical composition of red giant stars in Pal~4 --- one of the most remote GCs in the Milky Way --- and compare 
                 our abundance measurements to those for both low surface brightness dwarf galaxies, and GCs in the inner and the outer halo.} 
                 {By co-adding high-resolution, low-S/N Keck/HIRES spectra of 19 stars along the red giant branch, we estimate chemical abundance
                 ratios of 20 $\alpha$-, iron peak-, and neutron-capture elements. Our method gives total uncertainties on most element-to-iron ratios 
                 of typically 0.2 dex.}
                  {We measure ${\rm [Fe/H]} = -1.41\pm0.04~{\rm (statistical)} \pm0.17~{\rm (systematic)}$ and an $\alpha$-enhancement of [$\alpha$/Fe] = +0.38$\pm0.11$~dex, which is consistent with the canonical value of $\sim$ +0.4 dex found for Galactic halo field stars and most halo GCs at this 
                  metallicity. Although Pal~4 has higher enhancements in the heavier elements with respect to the halo, the majority of the element ratios are, within the measurement errors, consistent with those for  local halo field stars. We find, however, evidence for a lower [Mg/Ca] ratio than in other halo clusters.}
                   {Based on the available evidence, we conclude that the material from which Pal~4 and the Galactic halo formed experienced similar enrichment 
                   processes, despite the apparently younger age of this cluster. Within the limitations of our methodology we find no significant indication of an iron spread, as is typical of genuine GCs of the Milky Way. 
                   However, abundance ratios for individual stars in Pal~4 and other distant satellites are urgently needed to understand the relationship, if any, between remote GCs and other halo substructures (i.e., luminous and ultra-faint dwarf spheroidal galaxies).}

\maketitle 
%
%
%
%
%
%
\section{Introduction}
As the oldest readily identifiable stellar systems in the universe, globular clusters (GCs) are important tracers of the 
formation and early evolution of galaxies, the Milky Way (MW) included. Noting the apparent lack of   
a metallicity gradient among remote Galactic GCs, Searle \& Zinn (1978) proposed an accretion origin for the Galactic halo 
extending over a period of several Gyr.  Evidence for this picture of hierarchical halo growth has come from the existence of a
second-parameter problem among outer halo GCs (e.g., Catelan 2000; Dotter et al. 2010), which points to a significant age range
within this population. 

The remote GC Pal 4 is such an example of a second parameter cluster. At a Galactocentric distance of $R_ G = 109$ kpc 
(Stetson et al. 1999), it is one of only a few halo GCs at distances of  $\sim$100 kpc or beyond. With a half-light radius 
of $r_h \approx 23$ pc, it is also one of the most extended Galactic GCs currently known, being significantly larger 
than ``typical" GCs in the Milky Way or external galaxies (which have $\langle{r_h}\rangle \approx 3$~pc; see, 
e.g., Jord\'an et~al. 2005). In fact, with a total luminosity of just $L_V \sim 2.1\times10^4~L_{V,{\odot}}$, it is similar in
several respects to some of the more compact ``ultra-faint'' dwarf spheroidal (dSph) galaxies (Simon \& Geha 2007) 
that are being discovered  in the outer halo with increasing regularity (e.g., Belokurov et~al. 2007). 
Since almost nothing is known about their proper motions and internal dynamics, the relationship of faint, extended
GCs like Pal~4 to such low-luminosity galaxies is currently an open question. 

While there is a general consensus that Pal~4 is likely to be $\approx$ 1--2 Gyr younger than inner halo GCs of the same metallicity, 
such as M5, age estimates in the literature do not fully agree (e.g., Stetson et al. 1999 vs. Vandenberg 2000).
In particular, Stetson et al. (1999) note that an age difference with respect to the inner halo GCs could be explained if ``either 
[Fe/H] or [$\alpha$/Fe] for the outer halo clusters is significantly lower than ... assumed''. 
Conversely, Cohen \& Mel\'endez (2005a) found that the outer halo GC NGC~7492 (R$_{\rm GC}$ = 25 kpc) shows chemical abundance patterns 
that are very similar to inner halo GCs like M3 or M13. This similarity in the chemical enrichment now appears to extend into the outermost halo  
for at least some GCs: it has recently been shown that the abundance ratios in the remote (R$_{\rm GC}$ = 92 kpc) cluster Pal~3 
(Koch et al. 2009; hereafter Paper~I) bear a close resemblance to those of inner halo GCs.  
The chemical abundance patterns of remote halo GCs like Pal~3 and Pal~4 are important clues to the formation of the Milky Way, as they
allow for direct comparisons to those of the  dSph galaxies, which are widely believed to have been accreted into the halo
(e.g., Klypin et~al. 1999; Bullock et~al. 2001; Font et~al. 2006; Li et~al. 2009). 
In this spirit, Mackey \& Gilmore (2004) conclude that all young halo clusters (i.e., 30) did not originate in the MW but were 
donated by at least seven mergers with ``cluster-bearing'' dSph-type galaxies. 

There are, however, no high-dispersion abundance data yet published for this remote cluster. Previous low-resolution spectroscopic and 
photometric studies have 
established Pal~4 as a mildly metal-poor system, with [Fe/H] estimates ranging from $-1.28$ to $-1.7$ dex
(Armandroff et al. 1992; Stetson et al. 1999; Kraft \& Ivans 2003). 
In this paper, one of a series, we aim to extend the chemical element information for GCs in the Galactic halo out to the largest possible distances,
and to carry out a first analysis of Pal~4's chemical abundance patterns. 
As we have shown in Paper~I, which presented a similar analysis for Pal~3, it is possible to derive reliable abundance measurements for 
remote Galactic GCs by performing an integrated analysis of stacked, low signal-to-noise (S/N) --- but high-resolution --- spectra (see also McWilliam \& Bernstein 2008). 
Note, however, that this method presupposes that there is no significant abundance scatter present along the RGB and that all stars 
have the same mean abundances for all chemical elements. We have therefore no means of distinguishing Pal~4 as a genuine GC with 
no internal abundance spread from a dSph that may have a very broad abundance range (e.g., Shetrone et~al. 2001, 2003; Koch 2009), 
nor of discerning any intrinsic abundance variations (e.g., Lee et al. 2009).  We will return to this question in Section~5.2.
Neverthess, such studies can provide an important first step towards an overall characterization of the chemical element distribution,
and enrichment history, of the outer halo. 
\section{Data}
The spectra for Pal~4 were obtained during the same three nights in February and March 1999 as the spectra used in our analysis of Pal~3 (Paper I). 
During these observing runs, a total of 24 stars in Pal~4 were observed using the HIRES echelle spectrograph (Vogt et al. 1994) on 
the Keck I telescope.   Our spectroscopic targets were selected from a colour-magnitude diagram (CMD) constructed from $BV$ 
imaging obtained with the Low-Resolution Imaging Spectrometer (LRIS; Oke et al. 1995)
on the night of 13/14 January 1999. A CMD reaching roughly one magnitude below the main-sequence turnoff was constructed using a series of short and long 
exposures taken in both bandpasses (i.e., five exposures between 60s and 120s in $V$, and four exposures between 60s and 240s in $B$). 
\begin{figure}
\centering
\includegraphics[width=1\hsize]{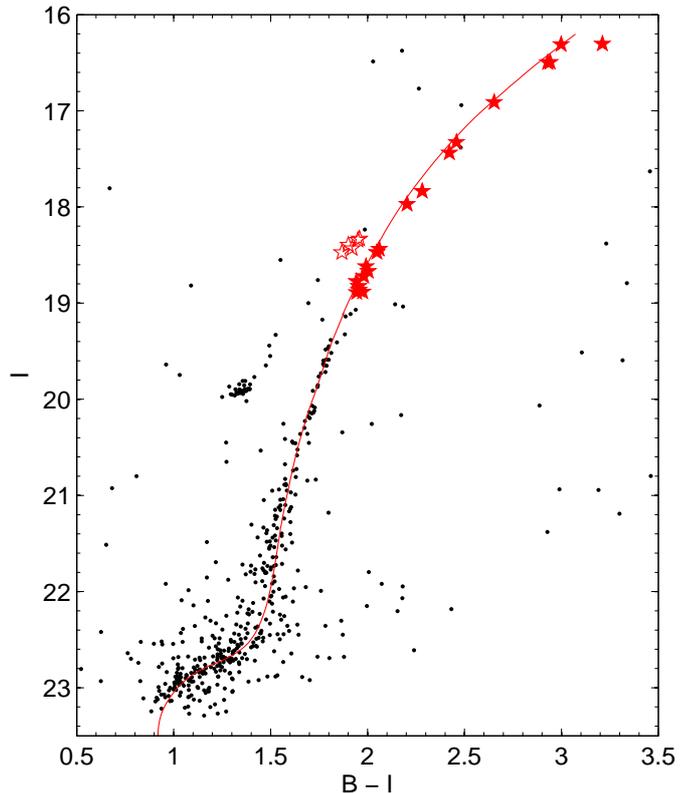}
\caption{CMD of Pal~4 based on photometry from Saha et al. (2005). The HIRES targets are highlighted as red symbols, with open stars denoting AGB candidates. 
Also shown 
is a scaled-solar Dartmouth isochrone (Dotter et al. 2008) with an age of 10 Gyr and a [Fe/H] of $-$1.4 dex, corrected for E(B$-V)$=0.01 and a distance modulus 
of 20.22 mag (Stetson et al. 1999).}
\end{figure}

Spectroscopic targets were identified from this CMD by selecting probable red 
giant branch (RGB) stars with $V \lesssim 20.25$. These stars all have cross identifications with the more recent work of Saha et al. (2005).  Fig.~1 shows the location of the target stars in the CMD from this latter work. 
As for Paper~I, we used a spectrograph setting that covers the wavelength range 
4450--6880 \AA\ with spectral gaps between adjacent orders, 
a slit width of 1.15$\arcsec$ and a CCD binning of 2$\times$2 in the spatial and spectral directions. This gives
a spectral resolution of $R\approx34000$.   
Each programme  star was observed for a total of 300--2400~s, depending on its apparent magnitude (see Table~1). 
\begin{table*}
\caption{Observation log and properties of the target stars}             
\centering          
\begin{tabular}{cccccccccc}     
\hline\hline       
 &  & Exposure time & $\alpha$  & $\delta$ & V & B$-$V & V$-$I & V$-$K & S/N  \\
\raisebox{1.5ex}[-1.5ex]{ID$^a$} & \raisebox{1.5ex}[-1.5ex]{Date} & [s] & (J2000.0) & (J2000.0)  & [mag] & [mag] & [mag] & [mag] & [pixel$^{-1}$] \\
\hline
Pal4-1  (S196) & Feb 11 1999, Mar 10 1999 & 3$\times$300  &                       11 29 17.13 & +28 57 59.9 & 17.81 & 1.46 & 1.52 &  3.43 & 8 \\
Pal4-2  (S169) & Feb 11 1999 & 2$\times$300  &  				  11 29 17.02 & +28 57 51.5 & 17.93 & 1.46 & 1.45 &  3.57 & 8 \\
Pal4-3  (S277) & Feb 11 1999 & 1$\times$300  &  				  11 29 13.24 & +28 58 13.6 & 17.82 & 1.66 & 1.53 &  3.69 & 8 \\
Pal4-5  (S434) & Feb 11 1999, Feb 12 1999 & 2$\times$300  & 			  11 29 16.67 & +28 58 42.1 & 17.95 & 1.44 & 1.47 &  2.81 & 7 \\
Pal4-6  (S158) & Feb 11 1999, Mar 10 1999 & 3$\times$420   & 			  11 29 15.50 & +28 57 47.0 & 18.22 & 1.30 & 1.33 &  3.06 & 7 \\
Pal4-7  (S381) & Feb 11 1999, Mar 10 1999 & 2$\times$600  & 			  11 29 14.83 & +28 58 32.2 & 18.55 & 1.19 & 1.24 &  2.79 & 7 \\
Pal4-8  (S364) & Feb 11 1999 & 1$\times$600  &  				  11 29 12.66 & +28 58 29.6 & 18.65 & 1.17 & 1.23 &  2.85 & 6 \\
Pal4-9  (S534) & Feb 11 1999 & 1$\times$750  &  				  11 29 13.32 & +28 59 07.6 & 19.00 & 1.08 & 1.18 &  3.62 & 6 \\
Pal4-10 (S325) & Feb 11 1999 & 2$\times$900  &  				  11 29 15.71 & +28 57 23.4 & 19.09 & 1.05 & 1.13 & \dots & 6 \\ 
Pal4-11 (S430)$^b$ & Feb 11 1999 & 1$\times$1200  & 				  11 29 13.82 & +28 58 40.9 & 19.35 & 0.89 & 1.04 & \dots & 5 \\
Pal4-12 (S328)$^b$ & Feb 11 1999, Mar 10 1999 & 1$\times$1200,1$\times$2400  & 	  11 29 17.63 & +28 58 25.1 & 19.35 & 0.90 & 1.03 & \dots & 8 \\
Pal4-15 (S307)$^b$ & Feb 11 1999 & 1$\times$1200  & 				  11 29 16.45 & +28 58 18.4 & 19.38 & 0.88 & 1.00 & \dots & 5 \\
Pal4-16 (S306)$^b$ & Feb 11 1999 & 1$\times$1200  & 				  11 29 17.77 & +28 58 19.5 & 19.43 & 0.88 & 1.02 & \dots & 5 \\
Pal4-17 (S472)$^b$ & Feb 12 1999 & 1$\times$1080  & 				  11 29 15.95 & +28 58 47.8 & 19.45 & 0.85 & 0.99 & \dots & 5 \\
Pal4-18 (S186) & Feb 11 1999 & 1$\times$1200  & 				  11 29 15.37 & +28 57 55.8 & 19.48 & 0.98 & 1.06 & \dots & 5 \\
Pal4-19 (S283) & Feb 12 1999 & 1$\times$1080  & 				  11 29 15.65 & +28 57 14.7 & 19.53 & 0.95 & 1.08 & \dots & 4 \\
Pal4-21 (S457) & Feb 11 1999 & 1$\times$1200  & 				  11 29 14.03 & +28 58 45.7 & 19.64 & 0.93 & 1.04 & \dots & 4 \\
Pal4-23 (S235) & Feb 12 1999 & 1$\times$1500  & 				  11 29 16.93 & +28 58 06.8 & 19.70 & 0.93 & 1.05 & \dots & 5 \\
Pal4-24 (S154) & Feb 11 1999 & 1$\times$1500  & 				  11 29 17.24 & +28 57 46.7 & 19.74 & 0.92 & 1.03 & \dots & 5 \\
Pal4-25 (S476) & Feb 12 1999 & 1$\times$1500  & 				  11 29 15.95 & +28 58 47.8 & 19.77 & 0.91 & 1.02 & \dots & 5 \\
Pal4-26 (S265) & Feb 12 1999 & 1$\times$1500  & 				  11 29 17.32 & +28 58 12.8 & 19.83 & 0.91 & 1.02 & \dots & 5 \\
Pal4-28 (S426) & Feb 12 1999 & 1$\times$1500  & 				  11 29 18.50 & +28 58 41.0 & 19.87 & 0.91 & 1.02 & \dots & 4 \\
Pal4-30 (S276) & Mar 10 1999 & 1$\times$1800  & 				  11 29 08.80 & +28 58 13.1 & 19.89 & 0.90 & 1.02 & \dots & 5 \\
Pal4-31 (S315) & Feb 12 1999 & 1$\times$1500  & 				  11 29 16.82 & +28 58 21.5 & 19.89 & 0.93 & 1.03 & \dots & 5 \\
\hline                                    
\end{tabular}
\\$^a$IDs preceded by ``S'' are cross-identifications from Table~7 of Saha et al. (2005).
\\$^b$Likely AGB stars.
\end{table*}
Table~1 also  lists the photometric properties of the target stars, where the $BVI$ photometry is taken from Saha et al. (2005) and the $K$-band magnitudes are from 2MASS (Skrutskie et al. 2006). 

The spectroscopic data were reduced using the MAKEE\footnote{MAKEE was developed by T. A. Barlow specifically for reduction of Keck HIRES data. 
It is freely available on the World Wide Web at the Keck Observatory home page, \tt{http://www2.keck.hawaii.edu/inst/hires/makeewww}} 
data reduction package. 
Because our spectra were obtained within a program to study the internal cluster dynamics (C\^ot\'e et al. 2002; Jordi et~al. 2009; Baumgardt et~al. 2009), 
the short exposure times --- which were chosen adaptively based on target magnitude --- lead to low signal-to-noise (S/N) ratios. 
Hence, the spectra are adequate for the measurement of accurate radial velocities but not for abundance measurements of individual stars.
For instance, we typically reach S/N ratios of 4--8 per pixel in the order containing H$\alpha$. 

Radial velocities of the individual targets were measured from a cross correlation against a synthetic red giant spectrum with 
stellar parameters representative of the Pal~4 target stars. The template covered the entire HIRES wavelength range, but excluded the spectral gaps. 
All our targets are consistent with the cluster's mean radial velocity of $\langle v_r \rangle = $ 72.9$\pm$0.3 km\,s$^{-1}$ (mean error) within the measurement errors
(see also Olszewski et~al. 1986). 
A detailed account of the dynamics of Pal~4 will be given in a separate paper. 

As in Paper~I,  we stack the individual spectra to increase their S/N ratio and to be able to perform an {\em integrated} abundance analysis 
(see also McWilliam \& Bernstein 2008).   In practice,  the spectra were Doppler-shifted and average-combined after weighting by their 
individual S/N ratios to yield a higher S/N spectrum which we can use to  place constraints on the chemical element abundances. 
As the CMD (Fig.~1) shows,  five of the stars appear to lie on the AGB (open symbols). We therefore constructed three different 
co-added spectra: i.e., using only the RGB stars, only the AGB stars, and the entire sample. 
The overall S/N ratios of the co-added spectra are (12, 25, 28) for the AGB, RGB, and the entire sample, respectively.  
A sample region of those spectra is shown in Fig.~2.
\begin{figure}
\centering
\includegraphics[width=1\hsize]{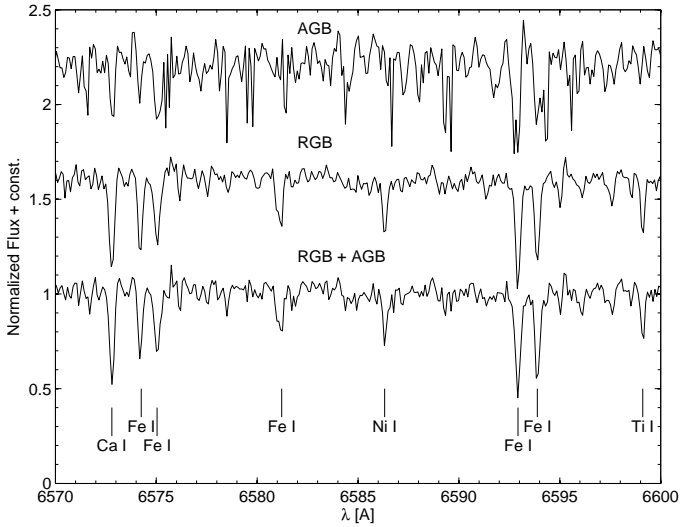}
\caption{A portion of the co-added spectra in one order with relatively high S/N ratio. A few absorption lines are designated. Also indicated is the subsample of stars that was included in the co-additions.}
\end{figure}
It is obvious from this figure that the pure AGB spectrum still has a too low S/N ratio for meaningful abundance measurements. Moreover, adding the AGB spectra to those of the higher-S/N spectra for the RGB stars may introduce additional noise to some features rather than improving the spectral quality. We therefore choose focus our abundance analysis on the co-added RGB sample only. 
\section{Abundance analysis}
As in our previous works (e.g., Paper~I),  we used model atmospheres interpolated from the updated grid of the 
Kurucz\footnote{\tt http://cfaku5.cfa.harvard.edu/grids.html} one-dimensional 72-layer, plane-parallel, line-blanketed models 
without convective overshoot and assuming local thermodynamic equilibrium (LTE) for all species. 
For this GC study, we used models that incorporate Castelli \& Kurucz's (
2003)\footnote{{\tt http://wwwuser.oat.ts.astro.it/castelli}} $\alpha$-enhanced opacity distribution functions, AODFNEW. 
This choice seems justified, since the majority of the metal-poor Galactic halo GCs, as well as the outer halo object Pal~3 (Paper~I),  are enhanced in the $\alpha$-elements by $\approx +0.4$ dex, so it seems plausible that Pal~4 will also follow this trend.   
Throughout this work, we used the 2002 version of the stellar abundance code MOOG (Sneden 1973) for all abundance calculations. We  
place our measurements on the Solar scale of Asplund et al. (2009).
\subsection{Linelist}
We derive chemical element abundances through standard equivalent width (EW) measurements that closely follow the procedures outlined in 
Paper~I. The main difference is that we are {\em only} dealing with an analysis of the co-added EWs in the present study, which thus requires an analogous 
treatment  of the synthetic EWs. 
The linelist for the present study is the same as already used in Paper~I and we refer the reader to that work for full details on the origin of the line data.
In practice, we measured EWs in the co-added spectra  (\S~2) by  fitting a Gaussian profile to the absorption lines using IRAF task {\em splot};  those value 
are recorded in Table~2. 
\onllongtab{2}{
\begin{longtable}{cccrr|cccrr}  
\caption{Linelist. ``HFS'' indicates that hyperfine splitting was taken into account for these transitions.} \\            
\hline
\hline
& $\lambda$ & E.P. &  & EW [m\AA] & & $\lambda$ & E.P. &  & EW [m\AA]  \\
\raisebox{1.5ex}[-1.5ex]{Element} & [\AA] & [eV]  & \raisebox{1.5ex}[-1.5ex]{log\,$gf$} & (RGB) & 
\raisebox{1.5ex}[-1.5ex]{Element} & [\AA] & [eV]  & \raisebox{1.5ex}[-1.5ex]{log\,$gf$} & (RGB) \\
\hline
\endfirsthead
\caption{Continued.} \\
\hline
& $\lambda$ & E.P. &  & EW [m\AA] & & $\lambda$ & E.P. &  & EW [m\AA]  \\
\raisebox{1.5ex}[-1.5ex]{Element} & [\AA] & [eV]  & \raisebox{1.5ex}[-1.5ex]{log\,$gf$} & (RGB) & 
\raisebox{1.5ex}[-1.5ex]{Element} & [\AA] & [eV]  & \raisebox{1.5ex}[-1.5ex]{log\,$gf$} & (RGB) \\
\hline                    
\endhead
\hline
\endfoot
\hline
\endlastfoot
Mg  I & 5528.42 &  4.35 & $-$0.357 & 177 & Cr  I & 5300.75 &  0.98 & $-$2.120 &  113 \\ 
Mg  I & 5711.09 &  4.33 & $-$1.728 & 108 & Cr  I & 5329.14 &  2.91 & $-$0.064 &   79 \\ 
Al  I & 6696.03 &  3.14 & $-$1.347 &  36 & Cr  I & 5345.81 &  1.00 & $-$0.980 &  165 \\ 
Si  I & 5684.48 &  4.95 & $-$1.650 &  33 & Cr  I & 5348.33 &  1.00 & $-$1.290 &  144 \\ 
Si  I & 5708.41 &  4.95 & $-$1.470 &  88 & Cr  I & 6330.09 &  0.94 & $-$2.914 &   50 \\ 
Si  I & 5948.55 &  5.08 & $-$1.230 &  64 & Mn  I$^{\rm HFS}$ & 5394.63 &  0.00 & $-$3.503 &  166 \\ 
Si  I & 6142.48 &  5.62 & $-$0.920 &  22 & Mn  I$^{\rm HFS}$ & 5432.51 &  0.00 & $-$3.800 &  136 \\ 
Si  I & 6155.13 &  5.61 & $-$0.750 &  64 & Mn  I$^{\rm HFS}$ & 6013.48 &  3.07 & $-$0.251 &  102 \\ 
Ca  I & 5261.71 &  2.52 & $-$0.580 & 107 & Mn  I$^{\rm HFS}$ & 6016.62 &  3.08 & $-$0.216 &  111 \\ 
Ca  I & 5590.13 &  2.52 & $-$0.570 & 114 & Mn  I$^{\rm HFS}$ & 6021.75 &  3.08 &	0.034 &   93 \\ 
Ca  I & 5601.29 &  2.53 & $-$0.520 & 116 & Fe  I & 4903.32 &  2.88 & $-$0.926 &  171 \\ 
Ca  I & 5857.46 &  2.93 &    0.230 & 157 & Fe  I & 4938.82 &  2.88 & $-$1.077 &  121 \\ 
Ca  I & 6166.44 &  2.52 & $-$1.140 & 103 & Fe  I & 4939.69 &  0.86 & $-$3.240 &  142 \\ 
Ca  I & 6169.04 &  2.52 & $-$0.800 & 126 & Fe  I & 5001.87 &  3.88 &	0.050 &  116 \\ 
Ca  I & 6169.56 &  2.52 & $-$0.480 & 143 & Fe  I & 5006.12 &  2.82 & $-$0.662 &  173 \\ 
Ca  I & 6455.60 &  2.52 & $-$1.290 &  95 & Fe  I & 5028.13 &  3.57 & $-$1.122 &   84 \\ 
Ca  I & 6471.67 &  2.52 & $-$0.875 & 122 & Fe  I & 5044.21 &  2.85 & $-$2.059 &  149 \\ 
Ca  I & 6499.65 &  2.52 & $-$0.820 & 115 & Fe  I & 5048.44 &  3.94 & $-$1.029 &  118 \\ 
Ca  I & 6717.69 &  2.71 & $-$0.610 & 136 & Fe  I & 5060.07 &  0.00 & $-$5.460 &  147 \\ 
Sc II & 5031.02 &  1.36 & $-$0.260 &  95 & Fe  I & 5068.77 &  2.94 & $-$1.041 &  159 \\ 
Sc II & 5239.81 &  1.46 & $-$0.770 &  85 & Fe  I & 5131.48 &  2.22 & $-$2.515 &  102 \\ 
Sc II & 5669.04 &  1.50 & $-$1.120 &  78 & Fe  I & 5145.10 &  2.20 & $-$2.876 &  106 \\ 
Sc II & 5684.19 &  1.51 & $-$1.050 &  77 & Fe  I & 5159.05 &  4.28 & $-$0.820 &   79 \\ 
Sc II & 6245.62 &  1.51 & $-$0.980 &  88 & Fe  I & 5162.28 &  4.18 &	0.020 &  157 \\ 
Sc II & 6604.60 &  1.36 & $-$1.480 &  36 & Fe  I & 5166.28 &  0.00 & $-$4.123 &  170 \\ 
Ti  I & 4997.10 &  0.00 & $-$1.722 & 132 & Fe  I & 5192.35 &  3.00 & $-$0.421 &  163 \\ 
Ti  I & 4999.51 &  0.83 &    0.140 & 180 & Fe  I & 5195.48 &  4.22 & $-$0.002 &  109 \\ 
Ti  I & 5001.01 &  2.00 & $-$0.052 &  78 & Fe  I & 5196.08 &  4.26 & $-$0.451 &   57 \\ 
Ti  I & 5009.65 &  0.02 & $-$1.900 & 105 & Fe  I & 5215.19 &  3.27 & $-$0.871 &  133 \\ 
Ti  I & 5039.96 &  0.02 & $-$1.170 & 134 & Fe  I & 5216.28 &  1.61 & $-$2.150 &  175 \\ 
Ti  I & 5064.65 &  0.05 & $-$0.985 & 152 & Fe  I & 5217.39 &  3.21 & $-$1.070 &  130 \\ 
Ti  I & 5147.48 &  0.00 & $-$1.876 & 149 & Fe  I & 5225.52 &  0.11 & $-$4.789 &  175 \\ 
Ti  I & 5152.19 &  0.02 & $-$1.912 & 134 & Fe  I & 5242.49 &  3.62 & $-$0.967 &  110 \\ 
Ti  I & 5173.75 &  0.00 & $-$1.120 & 178 & Fe  I & 5247.05 &  0.09 & $-$4.946 &  175 \\ 
Ti  I & 5219.70 &  0.02 & $-$1.980 & 137 & Fe  I & 5250.22 &  0.12 & $-$4.938 &  153 \\ 
Ti  I & 5866.46 &  1.07 & $-$0.840 & 133 & Fe  I & 5266.56 &  3.00 & $-$0.490 &  149 \\ 
Ti  I & 5922.12 &  1.05 & $-$1.470 &  91 & Fe  I & 5281.80 &  3.04 & $-$0.833 &  178 \\ 
Ti  I & 5965.83 &  1.88 & $-$0.410 & 120 & Fe  I & 5302.31 &  3.28 & $-$0.720 &  160 \\ 
Ti  I & 6064.63 &  1.05 & $-$1.970 &  65 & Fe  I & 5307.37 &  1.61 & $-$2.987 &  141 \\ 
Ti  I & 6126.22 &  1.07 & $-$1.420 &  90 & Fe  I & 5339.94 &  3.27 & $-$0.720 &  136 \\ 
Ti  I & 6258.10 &  1.44 & $-$0.355 & 105 & Fe  I & 5369.97 &  4.37 &	0.536 &  141 \\ 
Ti  I & 6556.08 &  1.46 & $-$0.943 &  57 & Fe  I & 5379.57 &  3.68 & $-$1.514 &   88 \\ 
Ti  I & 6743.13 &  0.90 & $-$1.630 &  90 & Fe  I & 5389.49 &  4.42 & $-$0.410 &  108 \\ 
Ti II & 5005.16 &  1.57 & $-$2.550 &  46 & Fe  I & 5393.18 &  3.24 & $-$0.715 &  147 \\ 
Ti II & 5013.68 &  1.58 & $-$1.935 & 119 & Fe  I & 5424.08 &  4.32 & 0.520 &  133 \\ 
Ti II & 5185.91 &  1.89 & $-$1.350 & 115 & Fe  I & 5569.63 &  3.42 & $-$0.500 &  138 \\ 
Ti II & 5226.55 &  1.57 & $-$1.300 & 173 & Fe  I & 5618.64 &  4.21 & $-$1.275 &   68 \\ 
Ti II & 5336.78 &  1.58 & $-$1.700 & 150 & Fe  I & 5753.12 &  4.26 & $-$0.688 &   83 \\ 
Ti II & 5396.23 &  1.58 & $-$2.925 &  36 & Fe  I & 5763.00 &  4.21 & $-$0.450 &   92 \\ 
Ti II & 5418.77 &  1.58 & $-$1.999 &  69 & Fe  I & 5862.36 &  4.55 & $-$0.058 &  108 \\ 
Ti II & 6606.95 &  2.06 & $-$2.790 &  50 & Fe  I & 5909.98 &  3.21 & $-$2.587 &  100 \\ 
 V  I & 6039.72 &  1.06 & $-$0.651 &  78 & Fe  I & 5916.25 &  2.45 & $-$2.834 &  124 \\ 
 V  I & 6081.44 &  1.05 & $-$0.578 &  52 & Fe  I & 5934.65 &  3.93 & $-$1.170 &   77 \\ 
 V  I & 6135.36 &  1.05 & $-$0.746 &  68 & Fe  I & 5956.71 &  0.86 & $-$4.605 &  146 \\ 
 V  I & 6243.10 &  0.30 & $-$0.978 & 128 & Fe  I & 5976.78 &  3.94 & $-$1.310 &   51 \\ 
 V  I & 6251.83 &  0.29 & $-$1.342 &  98 & Fe  I & 6024.06 &  4.55 & $-$0.120 &  106 \\ 
 V  I & 6274.66 &  0.27 & $-$1.670 &  66 & Fe  I & 6027.06 &  4.08 & $-$1.089 &   66 \\ 
Cr  I & 5247.57 &  0.96 & $-$1.640 & 146 & Fe  I & 6056.01 &  4.73 & $-$0.460 &   84 \\ 
Cr  I & 5296.70 &  0.98 & $-$1.400 & 146 & Fe  I & 6065.48 &  2.61 & $-$1.530 &  161 \\ 
Fe  I & 6078.49 &  4.79 & $-$0.424 &  81 & Fe II & 6432.68 &  2.89 & $-$3.708 &   23 \\ 
Fe  I & 6137.00 &  2.20 & $-$2.950 & 146 & Fe II & 6516.08 &  2.89 & $-$3.380 &   43 \\ 
Fe  I & 6173.34 &  2.22 & $-$2.880 & 155 & Co  I$^{\rm HFS}$ & 5301.01 &  1.71 & $-$2.000 &   72 \\ 
Fe  I & 6180.21 &  2.73 & $-$2.586 & 100 & Co  I$^{\rm HFS}$ & 5483.31 &  1.71 & $-$1.488 &  111 \\ 
Fe  I & 6213.44 &  2.22 & $-$2.481 & 142 & Co  I$^{\rm HFS}$ & 6814.89 &  1.96 & $-$1.900 &   86 \\ 
Fe  I & 6219.29 &  2.20 & $-$2.448 & 137 & Ni  I & 5035.36 &  3.63 &	0.290 &   96 \\ 
Fe  I & 6229.23 &  2.83 & $-$2.805 &  68 & Ni  I & 5080.53 &  3.65 &	0.134 &   75 \\ 
Fe  I & 6232.64 &  3.65 & $-$1.223 & 115 & Ni  I & 5084.09 &  3.68 &	0.034 &  111 \\ 
Fe  I & 6240.65 &  2.22 & $-$3.173 &  91 & Ni  I & 5146.48 &  3.71 & $-$0.060 &   70 \\ 
Fe  I & 6246.32 &  3.60 & $-$0.733 &  98 & Ni  I & 5578.71 &  1.68 & $-$2.641 &  130 \\ 
Fe  I & 6252.56 &  2.40 & $-$1.687 & 141 & Ni  I & 5587.85 &  1.94 & $-$2.142 &   96 \\ 
Fe  I & 6254.25 &  2.28 & $-$2.443 & 145 & Ni  I & 5592.26 &  1.95 & $-$2.588 &   80 \\ 
Fe  I & 6265.14 &  2.18 & $-$2.550 & 135 & Ni  I & 6128.97 &  1.68 & $-$3.390 &   73 \\ 
Fe  I & 6270.23 &  2.86 & $-$2.000 &  96 & Ni  I & 6176.82 &  4.09 & $-$0.430 &   52 \\ 
Fe  I & 6271.28 &  3.33 & $-$2.703 &  53 & Ni  I & 6177.25 &  1.83 & $-$3.600 &   48 \\ 
Fe  I & 6322.69 &  2.59 & $-$2.426 & 139 & Ni  I & 6327.59 &  1.68 & $-$3.090 &   97 \\ 
Fe  I & 6335.34 &  2.20 & $-$2.177 & 141 & Ni  I & 6378.26 &  4.15 & $-$0.820 &   40 \\ 
Fe  I & 6336.83 &  3.69 & $-$0.856 & 136 & Ni  I & 6482.81 &  1.94 & $-$2.630 &   84 \\ 
Fe  I & 6344.15 &  2.43 & $-$2.923 & 119 & Ni  I & 6586.32 &  1.95 & $-$2.812 &   91 \\ 
Fe  I & 6355.03 &  2.84 & $-$2.350 &  84 & Ni  I & 6767.78 &  1.83 & $-$2.170 &  121 \\ 
Fe  I & 6358.69 &  0.86 & $-$4.468 & 147 & Cu  I$^{\rm HFS}$ & 5105.51 &  1.39 & $-$1.505 &   90 \\ 
Fe  I & 6400.00 &  3.60 & $-$0.520 & 128 & Cu  I$^{\rm HFS}$ & 5782.06 &  1.64 & $-$1.720 &   79 \\ 
Fe  I & 6400.31 &  0.91 & $-$3.897 & 168 &  Y II & 4883.68 &  1.08 &	0.071 &  110 \\ 
Fe  I & 6475.63 &  2.56 & $-$2.941 & 103 &  Y II & 4900.11 &  1.03 & $-$0.090 &  115 \\ 
Fe  I & 6481.88 &  2.28 & $-$2.960 & 130 &  Y II & 5087.42 &  1.08 & $-$0.156 &   84 \\ 
Fe  I & 6498.95 &  0.96 & $-$4.687 & 144 &  Y II & 5200.41 &  0.99 & $-$0.570 &   98 \\ 
Fe  I & 6518.37 &  2.83 & $-$2.450 & 105 &  Y II & 5509.90 &  0.99 & $-$1.015 &   93 \\ 
Fe  I & 6574.22 &  0.99 & $-$5.004 & 115 & Zr II & 5112.28 &  1.66 & $-$0.590 &   42 \\ 
Fe  I & 6581.21 &  1.48 & $-$4.680 &  57 & Ba II & 4554.03 &  0.00 &	0.170 &  274 \\ 
Fe  I & 6609.12 &  2.56 & $-$2.692 & 128 & Ba II & 5853.00 &  0.60 & $-$1.010 &  130 \\ 
Fe  I & 6739.52 &  1.56 & $-$4.794 &  59 & Ba II & 6141.73 &  0.70 & $-$0.077 &  212 \\ 
Fe  I & 6750.15 &  2.42 & $-$2.608 & 140 & Ba II & 6496.91 &  0.60 & $-$0.380 &  201 \\ 
Fe II & 4923.93 &  2.89 & $-$1.307 & 166 & La II & 5114.56 &  0.23 & $-$1.060 &   55 \\ 
Fe II & 4993.35 &  2.81 & $-$3.485 &  37 & La II & 6390.46 &  0.32 & $-$1.400 &   45 \\ 
Fe II & 5197.58 &  3.23 & $-$2.233 &  88 & Ce II & 5274.23 &  1.04 &	0.150 &   33 \\ 
Fe II & 5234.63 &  3.22 & $-$2.220 &  94 & Nd II & 5249.59 &  0.98 &	0.217 &   54 \\ 
Fe II & 5425.26 &  3.20 & $-$3.372 &  22 & Dy II & 5169.69 &  0.10 & $-$1.660 &   12 \\ 
Fe II & 6247.56 &  3.89 & $-$2.329 &  24 &       &	   &	   &	      &      \\ 
\end{longtable}
}
Aided by the stellar atmospheres (described in detail in the next section), we computed theoretical EWs for the transitions in our line list using MOOG's {\em ewfind} driver 
and combined them into a mean value, $\langle{EW}\rangle$, using the same weighting scheme as for the observations,
\begin{equation} 
\langle{EW}\rangle = \, \frac{ \sum_{i=1}^N\,w_i\,EW_i\, }{ \,\sum_{i=1}^N\,w_i}, 
\end{equation}
where the weights $w_i$ are proportional to the S/N ratios as in the case of co-adding the observed spectra.  
The abundance ratio of each element was then varied until the predicted $\langle{EW}\rangle$ matched the observed EW for each line to yield
 the cluster's integrated chemical element ratio. 
Note that this method presupposes that there is no significant abundance scatter present along the RGB and all stars 
have the same mean abundances for all chemical elements. 
For the following analysis, we restricted the linelist to the more reliable features with EW$<$180 m\AA.  
For a few cases, such as Al, Zr, Ce, and Dy, the stated abundance ratios are based on marginal detections of only one line with usually about 
30--40 m\AA~widths. 
Unfortunately, neither of the important elements O and Eu could be detected: while the stronger [O~I] 6300 \AA~and Eu~II 6645 \AA~lines fall on the gap between the HIRES CCDs, the weaker 6363 \AA~(O) and 6437 \AA~(Eu) lines are strongly affected by telluric blends and spectral noise, which renders them unusable for the present work. Likewise, the Na-D lines are too strong to be reliably used in our analysis, while the only other transition covered by our spectra, the Na~I 5688 \AA~line, is too strongly affected by the low S/N ratio around that feature. 

We accounted for the effects of hyperfine structure for the stronger lines of the odd-Z elements Mn~I, Co~I, and Cu~I by extracting the predicted 
EW from MOOG's {\em blends} driver and using atomic data for the splitting from McWilliam et al. (1995). 
The effect on all other elements (such as Ba~II or La~II) was found to be typically less than 0.03 dex and thus 
much smaller than the usual systematic errors (\S~4) so that we ignored hyperfine splitting for all other elements. 
\subsection{Stellar parameters}
We derived effective temperatures (T$_{\rm eff}$) for each star using its photometry, in particular, 
from B$-$V and V$-$I colors using the data from Saha et al. (2005). This information was supplemented with 
2MASS K-band photometry (Skrutskie et al. 2006) to obtain V$-$K
estimates  for the eight brightest stars (see Table~1).
We assumed a reddening of E(B$-$V)=0.01 (Stetson et al. 1999) with the 
extinction law of Winkler (1997). 
In practice, the  T$_{\rm eff}$-color calibrations of Ram\'{\i}rez \& Mel\'endez (2005) were applied for V$-$I and V$-$K, and the Alonso et al. (1999) transformations 
for B$-$V. All three values agree well with offsets of 6 and 20 K for B$-$V vs. V$-$I and V$-$K, with an $rms$ scatter of 50 and 80~K, respectively. 
For all these calibrations, we adopted the cluster mean metallicity of $-$1.43 dex on the Kraft \& Ivans (2003) scale. 
The resulting temperatures have a formal mean random error due to color and calibration uncertainties of 136 K  on average. 
In practice, we adopt an error-weighted mean of all three color indicators as the final T$_{\rm eff}$ for the atmospheres. 
Fig.~3 shows the distribution of effective temperatures for our targets.  
\begin{figure}[htb]
\centering
\includegraphics[width=1\hsize]{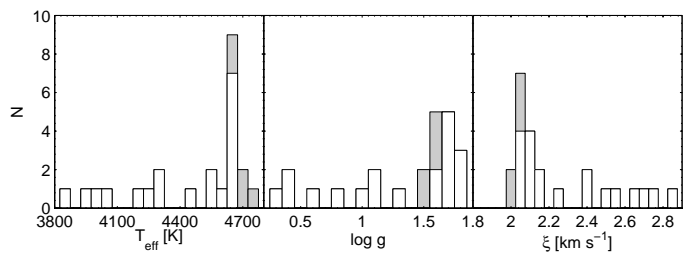}
\caption{Distribution of stellar parameters for the RBG and AGB stars (open and shaded histograms, respectively).}
\end{figure}

Surface gravities, log\,$g$, were derived from the  photometry, an adopted distance modulus of 20.22 mag (Stetson et al. 1999), 
and the above temperature and metallicity estimates. 
A mass of 0.85 M$_{\odot}$ was adopted for the red giants, 
as indicated by a comparison with the Dartmouth isochrones (Dotter et al. 2008; Fig.~1). 
Errors on the input parameters (predominantly that on T$_{\rm eff}$) lead to a typical uncertainty in log\,$g$ of  $\pm$0.16 dex. 
As in Paper~I, we derived microturbulent velocities, $\xi$, from a linear fit to the temperatures of halo stars  that have similar parameters to ours (Cayrel et~al. 2004). 
The scatter around the best-fit relation implies a typical error of $\sigma(\xi)$ $\approx$ 0.25 km\,s$^{-1}$. 

Since we have no prior knowledge of the individual stellar metallicities, we adopt the value of $-1.43$ dex (Kraft \& Ivans 2003) as representative of the cluster mean and as an input metallicity for the atmospheres. This value is then refined iteratively using the Fe I abundance from the previous step as input for the following atmosphere calculations. 

In addition, we calculate an independent metallicity estimate for individual stars from the Mg I line index at 5167, 5173 \AA, which is defined and calibrated on the scale of Carretta \& Gratton (1997) as in Walker et al. (2007) and Eq.~2 in Paper~I. For this, we assume a horizontal branch magnitude, V$_{\rm HB}$, of 20.8 mag (Stetson et al. 1999). 
Although we list the Mg~I indicator in our final abundance in Table~3, we emphasize that this value is meant as an initial estimate of the cluster metallicity, rather than a reliable measurement of its abundance scale.
Table~3 lists the final abundance ratios derived from the co-added red giant sample. Here, neutral species are given relative to Fe I, while 
the ratios of  ionized species are listed with respect to the ionized iron abundance as [X~II / Fe II].
\begin{table}
\caption{Abundance results from the co-added red giant spectrum}             
\centering          
\begin{tabular}{crcrc}     
\hline\hline       
Element &  [X/Fe] & $\sigma$ & N & $\sigma_{\rm tot}$ \\
\hline                    
Fe$_{\rm Mg I}^a$ & $-1.41$  & 0.28 & \dots  & \dots \\ 
Fe I  & $-$1.41 &  0.35 & 81 & 0.17  \\
Fe II & $-$1.54 &  0.21 &  8 & 0.25  \\
Mg I  &    0.25 &  0.21 &  2 & 0.20  \\
Al I  &    0.36 &  \dots &  1 & 0.19  \\
Si I  &    0.47 &  0.32 &  5 & 0.15  \\
Ca I  &    0.40 &  0.16 & 11 & 0.21  \\
Sc II &    0.29 &  0.28 &  6 & 0.14  \\
Ti I  &    0.24 &  0.32 & 18 & 0.30  \\
Ti II &    0.61 &  0.39 &  8 & 0.17  \\
 V I  &    0.16 &  0.17 &  6 & 0.29  \\
Cr I  & $-$0.18 &  0.19 &  7 & 0.28  \\
Mn I  & $-$0.18 &  0.20 &  5 & 0.23  \\
Co I  &    0.38 &  0.07 &  3 & 0.18  \\
Ni I  &    0.04 &  0.29 & 16 & 0.14  \\
Cu I  & $-$0.66 &  0.16 &  2 & 0.18  \\
 Y II &    0.30 &  0.29 &  5 & 0.17  \\
Zr II &    0.53 &  \dots &  1 & 0.16  \\
Ba II &    0.36 &  0.16 &  4 & 0.19  \\
La II &    0.67 &  0.10 &  2 & 0.12  \\
Ce II &    0.34 &  \dots &  1 & 0.16  \\
Nd II &    0.45 &  \dots &  1 & 0.16  \\
Dy II &    0.32 &  \dots &  1 & 0.16  \\   
\hline                  		       
\end{tabular}				       
\\$^a$Metallicity estimate based on the Mg I calibration of Walker 
et al. (2007), on the metallicity scale of Carretta \& Gratton (1997).
\end{table}
\section{Abundance errors}
As a measure of the random uncertainties on our abundance ratios, Table~3 also lists the 1$\sigma$-scatter of the line-by-line measurements together with the number of transitions, $N$, used in the analysis. This contribution is generally small for those species with many suitable transitions (e.g., Fe~I, Ca, Ti~I, Ni) yet dominates for the other, poorly-sampled  
elements. As in Paper~I, we adopt in what follows a minimum random abundance error of 0.10 dex 
and assign an uncertainty of 0.15 dex if only one line could be measured. 

In order to investigate the extent to which inaccurate radial velocity measurements can lead to a broadening of the observed lines during the co-addition 
of the individual, Doppler-shifted spectra, we carried out a series of 1000 Monte Carlo simulations. In each simulation, we corrected every spectrum by a velocity that accounted 
for the velocity error before combining those falsified spectra into a new spectrum. The EWs for the entire line list were then re-measured from each of those spectra 
in an automated manner. 
As a result, the EWs changed by (10$\pm$5)\% on average, with 1$\sigma$ of the widths changing by less than 15\%. 
We then repeated our abundance determinations by Monte Carlo varying the EWs by this amount and deriving new means and dispersions. This 
revealed that a 15\% uncertainty in the measured EW incurs an error of 0.04 dex on the mean iron abundance. For this, we conclude that inaccurate Doppler-shifts
of the spectra are not a major source of uncertainty in an analysis of this sort. The main contributor to the random errors are instead the EW measurements 
at these still-low S/N levels and, to a lesser extent, the standard uncertainties in the atmosphere models and atomic parameters themselves. 

Although none of our stars is a likely non-member in terms of our CMD selection, nor indicated by deviating gravity sensitive features as the Mg~b triplet or the Na-D lines, nor by  
discrepant radial velocity, we explored the effect of co-adding undesired foreground dwarfs to the red giant sample on the resulting abundance ratios. To this end, we computed synthetic spectra for each star, using the atmospheric parameters determined above and adopting the element ratios listed in Table~3. 
We then synthesized a spectrum of a metal-poor dwarf star (T$_{\rm eff}$ = 5700~K, log\,$g$ = 4.2, and $\xi$ = 1.1 km\,s$^{-1}$) 
and randomly replaced one or two of the RGB stars with a dwarf spectrum in the co-addition. The EWs of the resulting, co-added synthetic spectrum were then re-measured as above. 
As a consequence, the presence of one (two) underlying dwarf stars in the co-added spectrum does not change the co-added EWs by more than 5\% (9\%), on average.  
Thus our abundance ratios are insensitive to any residual foreground contamination, with no expected effect larger than 0.02 dex.   

Systematic uncertainties of the stellar parameters were evaluated from a standard error analysis (e.g., Koch \& McWilliam 2010). To this end, each parameter was 
varied by the typical uncertainty (T$_{\rm eff}\pm$150 K; log\,$g\pm$0.2 dex; $\xi\pm$0.25 km\,s$^{-1}$; see previous section), from which  new atmospheres were interpolated for each star. This assumes that all stars are systematically affected in the same manner by the same absolute error. Furthermore, the column labeled ``ODF'' shows the changes induced by using the Solar-scaled opacity distributions ODFNEW, which corresponds to an error in the $\alpha$-enhancement of 0.4 dex. Using these changed atmospheres, theoretical EWs were computed for each star and then combined into a new $\langle{EW}\rangle$ to be compared with the observed EW as before. We list in Table~4 the deviations of the resulting new abundances from the nominal values, [X/Fe], obtained from the unchanged atmospheres. 
Overall, the largest effect is naturally found with regard to T$_{\rm eff}$ errors, while changes in log\,$g$ mostly affect the ionized species (see also Paper~I). 
\begin{table}
\caption{Error analysis: deviations from the abundances in Table~3}
\centering          
\begin{tabular}{ccccr}
\hline
\hline
& $\Delta$T$_{\rm eff}$ & $\Delta\,\log\,g$ & $\Delta\xi$ &   \\
\raisebox{1.5ex}[-1.5ex]{Ion}  & $\pm$150\,K  & $\pm$0.2\,dex & $\pm$0.25\,km\,s$^{-1}$ & \raisebox{1.5ex}[-1.5ex]{ODF} \\
\hline
Fe I  & $\pm$0.13 & $\pm$0.01 & $\mp$0.12 & $-$0.02 \\
Fe II & $\mp$0.20 & $\pm$0.12 & $\mp$0.09 & $-$0.13 \\
Mg I  & $\pm$0.10 & $\mp$0.02 & $\mp$0.10 &  0.02  \\
Al I  & $\pm$0.13 & $\mp$0.02 & $\mp$0.01 & 0.04 \\
Si I  & $\mp$0.03 & $\pm$0.03 & $\mp$0.03 & $-$0.02 \\
Ca I  & $\pm$0.19 & $\mp$0.03 & $\mp$0.08 & 0.03 \\
Sc II & $\mp$0.02 & $\pm$0.07 & $\mp$0.05 & 0.04 \\
Ti I  & $\pm$0.30 & $\mp$0.03 & $\mp$0.08 & 0.01 \\
Ti II & $\mp$0.03 & $\pm$0.07 & $\mp$0.08 & 0.04 \\
 V I  & $\pm$0.32 & $\mp$0.02 & $\mp$0.03 & 0.00 \\
Cr I  & $\pm$0.26 & $\mp$0.03 & $\mp$0.10 & 0.02 \\
Mn I  & $\pm$0.22 & $\pm$0.01 & $\pm$0.04 &  0.04 \\
Co I  & $\pm$0.16 & $\pm$0.03 & $\pm$0.03 & 0.01 \\
Ni I  & $\pm$0.11 & $\pm$0.01 & $\mp$0.06 & $-$0.01 \\
Cu I  & $\pm$0.15 & $\pm$0.02 & $\mp$0.03 & 0.01 \\
 Y II & $\pm$0.01 & $\pm$0.06 & $\mp$0.10 & 0.04 \\
Zr II & $\mp$0.02 & $\pm$0.07 & $\mp$0.02 & 0.05 \\
Ba II & $\pm$0.05 & $\pm$0.06 & $\mp$0.15 & 0.01 \\
La II & $\pm$0.04 & $\pm$0.07 & $\mp$0.02 & 0.03 \\
Ce II & $\pm$0.02 & $\pm$0.07 & $\mp$0.01 & 0.04 \\
Nd II & $\pm$0.02 & $\pm$0.06 & $\mp$0.03 & 0.04 \\
Dy II & $\pm$0.05 & $\pm$0.06 & $\pm$0.00 & 0.04 \\
\hline                  
\end{tabular}
\end{table}

Finally, we interpolated the values in Table~4 to the actual parameter uncertainties estimated in Sect.~3.2 and adopted an error of the atmosphere $\alpha$-enhancement 
of $\pm$0.2 (in accordance with the results for [$\alpha$/Fe] in Table~3). 
These contributions were added in quadrature to the random error to yield the total abundance error, which we list as $\sigma_{\rm tot}$ in the last column of Table~3 and which we will show in the following figures unless noted otherwise. Since this procedure neglects the covariances between the stellar parameters, these errors can be regarded as {\it upper limits} on the actual abundance uncertainties.  
In the end, our measurements yield element ratios that are typically accurate to within 0.2 dex for the $\alpha$-elements, 0.15--0.30 dex for the iron peak elements, and
approximately 0.2 dex for the heavy elements. Although these error estimates may seem relatively large (and dominated by the systematic uncertainties), we have shown in Paper~I that 
the results from a co-added abundance analysis of this kind are largely consistent with those obtained from individual, high-S/N spectroscopic measurements. Thus, the present data are adequate for placing useful limits on the chemical abundances in Pal~4 and characterizing the general trends (see also Shetrone et al. 2009). 
 \section{Abundance results}
Our abundance measurements based on the co-added RGB spectrum are plotted in Fig.~4. Note that the values for [Al, Zr, Ce, Dy/Fe] are only upper limits (\S~3.1), although we show their formal, total error bars in this figure (cf. Fig.~9).
\begin{figure}[htb]
\centering
\includegraphics[width=1\hsize]{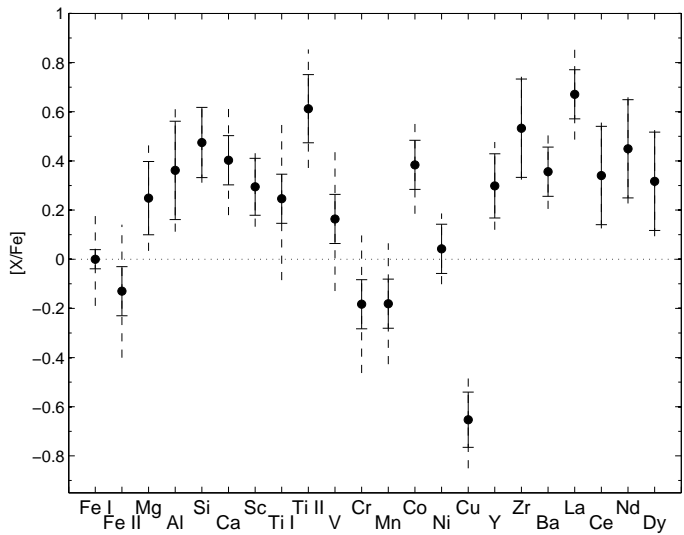}
\caption{Abundances ratios from the co-added RGB star spectrum. The dashed error bars indicate the total uncertainties 
(Tables~3 in 4), while the solid symbols represent 1$\sigma$ random errors.}
\end{figure}

\subsection{Iron}
Based on our sample of 19 RGB stars, we find a mean iron abundance of 
$${\rm [Fe I/H]} = -1.41\pm0.04~{\rm (statistical)} \pm0.17~{\rm (systematic)}.$$ 
This value is in excellent agreement with the Fe II based abundance scale of Kraft \& Ivans (2003), and slightly more metal-poor than the 
value of $-1.28\pm0.20$ dex reported by Armandroff et al. (1992) from the calcium triplet on the Zinn \& West (1984) scale, and by 
Stetson et al. (1999) from photometry. 
It is interesting to note that also the mean [Fe/H]$_{\rm Mg~I}$ from the Mg\,b index (\S~3.2) agrees very well with the Fe~I scale: for the red giant sample we find the same mean of $-1.41$ dex, with a 1$\sigma$ spread of 0.28 dex. 

Ionization equilibrium is not fulfilled in this integrated analysis to within the {\em random} uncertainties, while both stages agree if one accounts for their total errors; the mean deviation of the neutral and ionized species is [Fe\,{\sc i}/Fe\,{\sc ii}]=$0.13\pm$0.08 dex.  A similar deviation was found in an identical analysis of co-addded RGB star spectra  in Paper~I, although in the opposite sense  (i.e., with Fe I yielding higher abundances). 
As in Paper~I, we conclude that Fe II lines in general seem ill suited to establishing a population's iron abundance from a low-S/N spectral co-addition (cf. Kraft \& Ivans 2003; McWilliam \& Bernstein 2008). Typical EWs of the eight Fe II lines we used in the analysis fall in the range 20--90 m\AA. 
As Table~4 indicates, a systematic increase of 0.24 dex in the surface gravity would settle the ionization equilibrium at [Fe/H] of $-$1.39 dex, which is 
entirely consistent with the value found above from the neutral species. 
Moreover, a change in the temperature scale of just $-$54 K (without altering log\,$g$) would re-install the equilibrium at $-$1.44 dex (see also Koch \& McWilliam 2010).
In what follows, we therefore proceed with our adopted log\,$g$ scale and take the imbalance between ionized and neutral species at face value. 
\subsection{Tests for abundance spreads}
As argued earlier, an integrated abundance analysis works reliably under the {\em ad-hoc} assumption of the same chemical abundance for all stars that enter the 
co-added spectrum. Here we discuss several tests of how realistic this assumption is for our analysis of Pal~4. 

As a first test, we consider the spread in colour about the 
fiducial isochrone shown in Fig.~1. By interpolating a finely spaced isochrone grid in metallicity and using the identical values for age, distance modulus, and reddening 
as above, we find that the colour range of the RGB targets translates into a metallicity spread of 0.036 dex. Accounting for photometric errors, which propagate to a mean metallicity error of 
0.026 dex, we find an intrinsic spread of 0.025 dex in the photometric metallicities. Since this procedure did not include errors on the distance modulus or reddening, and 
uncertainties in the adopted age and $\alpha$-enhancement of the isochrones will lead to even larger uncertainties, we conclude that there is no evidence of any global abundance spread on the RGB, based on the photometric metallicities alone. This notion is consistent with the homogeneity (in iron or overall metallicity) of most genuine Galactic GCs  (e.g., Carretta et al. 2009).

Secondly, we divided the RGB sample in two halves and co-added spectra for each subsample\footnote{In practice stars were chosen to alternate in magnitude so that sample \#1 includes Pal4-1,3,6,8,10,19,23,25,28,31, and the remainder constitutes sample \#2.}. The above procedures to obtain iron abundance constraints from the co-added EWs were repeated and we find slightly more metal poor values for either subsample: $-1.46\pm0.05$ and $-1.45\pm0.06$ dex, respectively, where the stated uncertainties account for  random errors only. 
Therefore, there is no evidence of an abundance difference between the subsets within the measurement errors. 
Strictly, one would need to repeat this exercise for all (92378) possible combinations in order to detect the maximum abundance difference, which could  be indicative of any real spread. The measurement of the 81 Fe lines in this amount of spectra is, however, computational 
expensive and beyond our present scope. 

As a last test, we employed a line-coaddition technique within the spectrum of each individual star, similar to that outlined in Norris et al. (2007; and references therein); see also Koch et al. (2008c): For each star, the useable 81 Fe lines were thus shifted to zero wavelength at each line center, and then co-added into a composite, ``master line''. The same was carried out for a synthetic spectrum that matches the stellar parameters of the stars. This way we find a 1$\sigma$ dispersion of the 19 [Fe/H] values of 0.176 dex. If we account for the random measurement errors from this procedure and assume the same systematic uncertainties as in our proper analysis (Sect.~4), we estimate an intrinsic abundance spread of no more than 0.05 dex. This is most likely an upper limit, since radial velocity uncertainties may have a larger impact on this method, and it is also not self evident that the systematic errors are identical to those in Table~3. At this low internal dispersion, however, Pal~4 does not comply with the broad ranges found in the dSphs (e.g., Table~1 in Koch 2009), while it is consistent with the upper limit for GC homogeneity found in Carretta et al. (2009). We conclude that, within the limitations of our spectral co-addition techniques, Pal~4 most likely shows little to no abundance spread, rendering it a genuine (MW) GC and arguing against an origin in a dSph-like environment.
\subsection{Alpha-elements}
All $\alpha$-elements measured in this study are enhanced with respect to Fe. While the [Ca/Fe] and [Si/Fe] ratios show the canonical value of $\ga$ 0.4~dex typical for 
Galactic halo field and GC stars, the abundance ratios of Mg and Ti are slightly lower, at about 0.25~dex. Because the latter species have slightly larger errors, the error-weighted mean of all four elements is 
$${\rm {[}}\alpha/{\rm Fe]}  = 0.38\pm0.11~{\rm dex}.$$
 The $\alpha$-element ratios are shown for Mg, Ca, and Ti in Fig.~5 where they are compared to Galactic halo and disk data from the literature (small black dots). The data shown here are  taken from the same sources as in Paper~I. At this point, we draw the reader's attention to an important caveat in Fig.~5 and subsequent figures: the selection of halo stars used in these comparisons is, by necessity, a local sample. How appropriate it is to use local halo field stars in a comparison to remote halo GCs  is unclear, particularly if there are radial gradients in the abundance ratios, as has sometimes been claimed (e.g., Nissen \& Schuster 1997; Fulbright 2002).
We shall return in \S~5.6 to the issue of $\alpha$-element enhancements amongst different populations in the Galactic halo. 
\begin{figure}[htb]
\centering
\includegraphics[width=1\hsize]{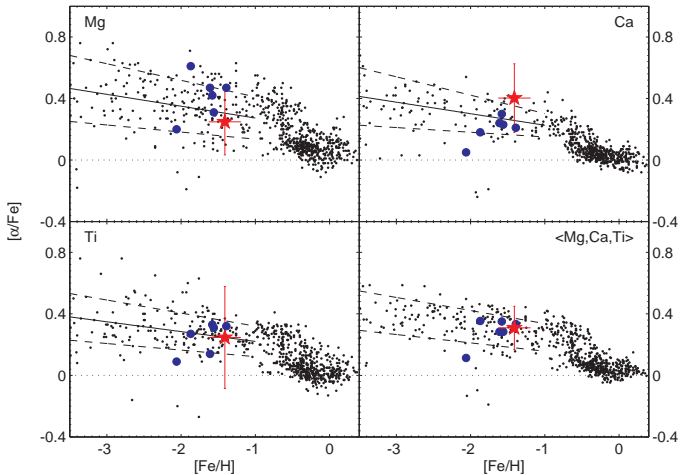}
\caption{[$\alpha$/Fe] abundance ratios for Pal~4 from this work (filled red star) in comparison with the GCs listed in Table~5 (solid dark blue circles). The solid and dashed lines illustrate the mean star relation and its $\pm$1$\sigma$ spread, respectively, from linear fits to the halo star data (black dots). See text for details.}
\end{figure}

We note in passing that, although the difference [Ti I/Ti II] = $-$0.24 dex is large, ionization equilibrium for Ti is satisfied considering the large combined total error for both species. This discrepancy is only significant at the 0.6$\sigma$-level and is in the 
opposite sense of the deviation in Fe. In any case, a detailed interpretation of any imbalances in terms of cumulative non-LTE effects 
along the RGB in our integrated abundance analysis would be beyond the scope of the present work (e.g., Koch \& McWilliam 2010).

At $-0.15$ dex, the [Mg/Ca] ratio is comparably low. While Mg is produced during the hydrostatic burning phases in the type II supernova (SN) progenitors, Ca nucleosynthesis proceeds during the SN explosion itself (e.g., Woosley \& Weaver 1995). Thus, it is not evident that one element should trace the other over a broad metallicity range. In fact, theoretical yields predict a delicate mass dependance of the [Mg/Ca] ratio. In Fig.~6, we show the distributions of this ratio for Galactic halo stars (gray shaded histogram) using the data of Gratton \& Sneden (1988; 1994), McWilliam  et~al. (1995),  Ryan et~al. (1996), 
Nissen \& Schuster (1997), McWilliam (1998),  Hanson et al. (1998), Burris et~al. (2000),  Fulbright (2000, 2002), 
Stephens \& Boesgaard (2002), Johnson (2002),  Ivans et~al. (2003) and Cayrel et~al. (2004). 

Fig.~6 also shows the currently available measurements for Local Group dSph galaxies (black line in Fig.~6) by Shetrone et~al. (2001; 2003; 2009), Sadakane et~al. (2004), Monaco et~al. (2005), Letarte (2007), Koch et~al. (2008a,b), Frebel et~al. (2010), Aoki et~al. (2009), Cohen \& Huang (2009) and Feltzing et al. (2009); see also Koch (2009).  Halo stars scatter around a [Mg/Ca] of zero, with mean and 1$\sigma$ dispersion of 0.05 and 0.15 dex, respectively. Stars
with very low abundance ratios are the exception (e.g., Lai et~al. 2009). In fact, the third moment of the halo distribution, at +0.55, indicates a higher-[Mg/Ca] tail. The dSph galaxies, on the other hand, have a formal mean and dispersion of 0.12 and 0.23 dex.  It is important to bear in mind, though, 
that the abundance ratios in the dSphs are inevitably unique characteristics of each galaxy and should be governed by their individual star formation 
histories and global properties (e.g., Lanfranchi \& Matteucci 2004). In particular the so-called ultra-faint dSph galaxies, which have very low masses, show a propensity  to reach higher [Mg/Ca] ratios as a result of a stochastical sampling of the high-mass end of the IMF, which in turn causes an imbalance between the Mg- and Ca-production (e.g., Koch et al. 2008a; Feltzing et al. 2009; Norris et al. 2010). In addition, the dSph galaxies show a clear extension towards low [Mg/Ca] ratios, which reflects in an overall skewness of $-$0.13 in the dSph distribution. Notably, all of the ``reference GCs" considered here (Table~5) have  positive Mg/Ca values. 
Given the rather large formal uncertainty of $\pm$0.30 dex on the [Mg/Ca] ratio (adding the total errors on Mg and Ca/Fe in quadrature) our measurement
does not serve as an especially strong discriminator between halo field or dSph origin for Pal~4. Nevertheless, its value is clearly different from those of the remainder of inner and outer halo GCs, and may point to different enrichment processes in the environment where Pal~4 formed. 

\begin{figure}[htb]
\centering
\includegraphics[width=1\hsize]{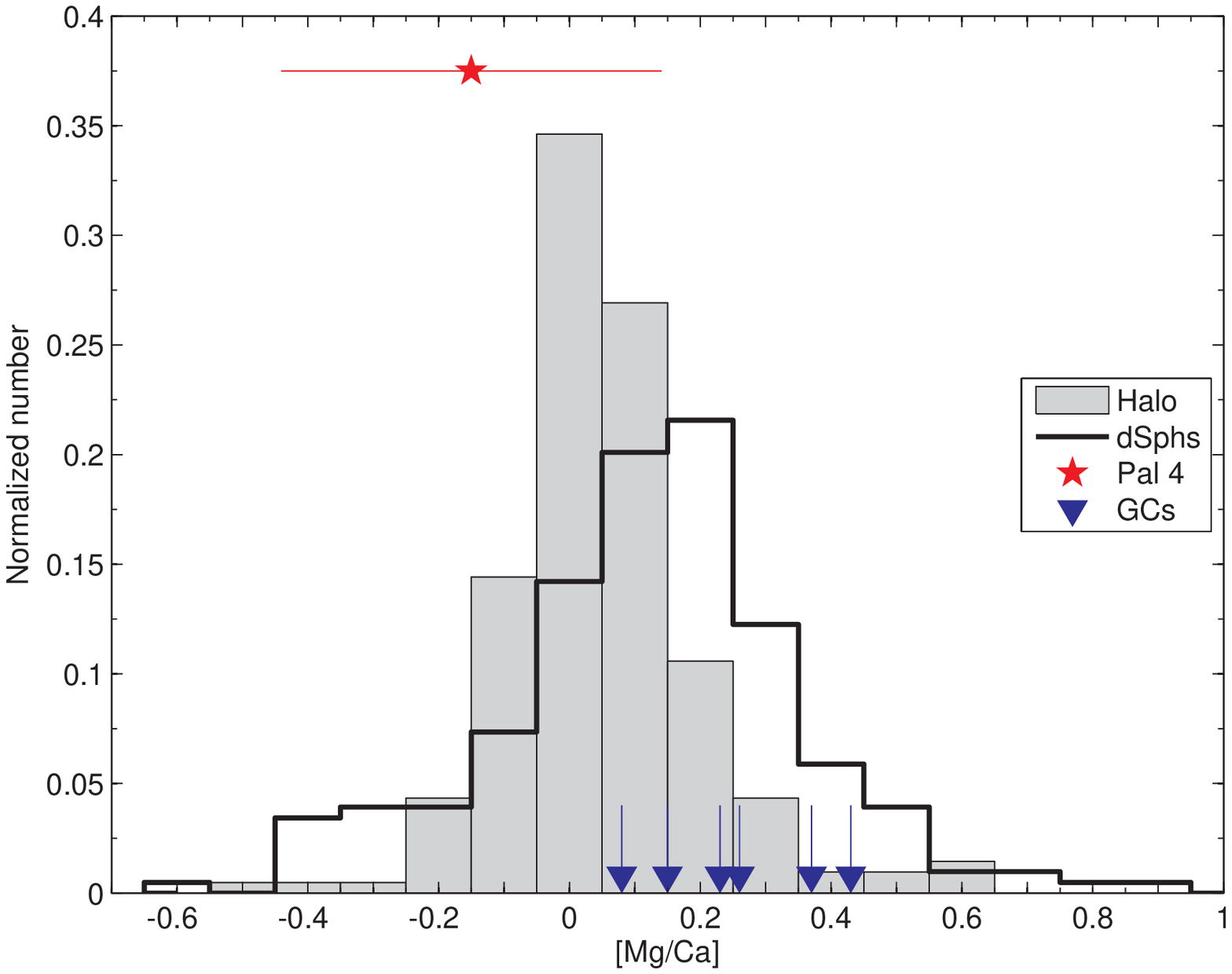}
\includegraphics[width=1\hsize]{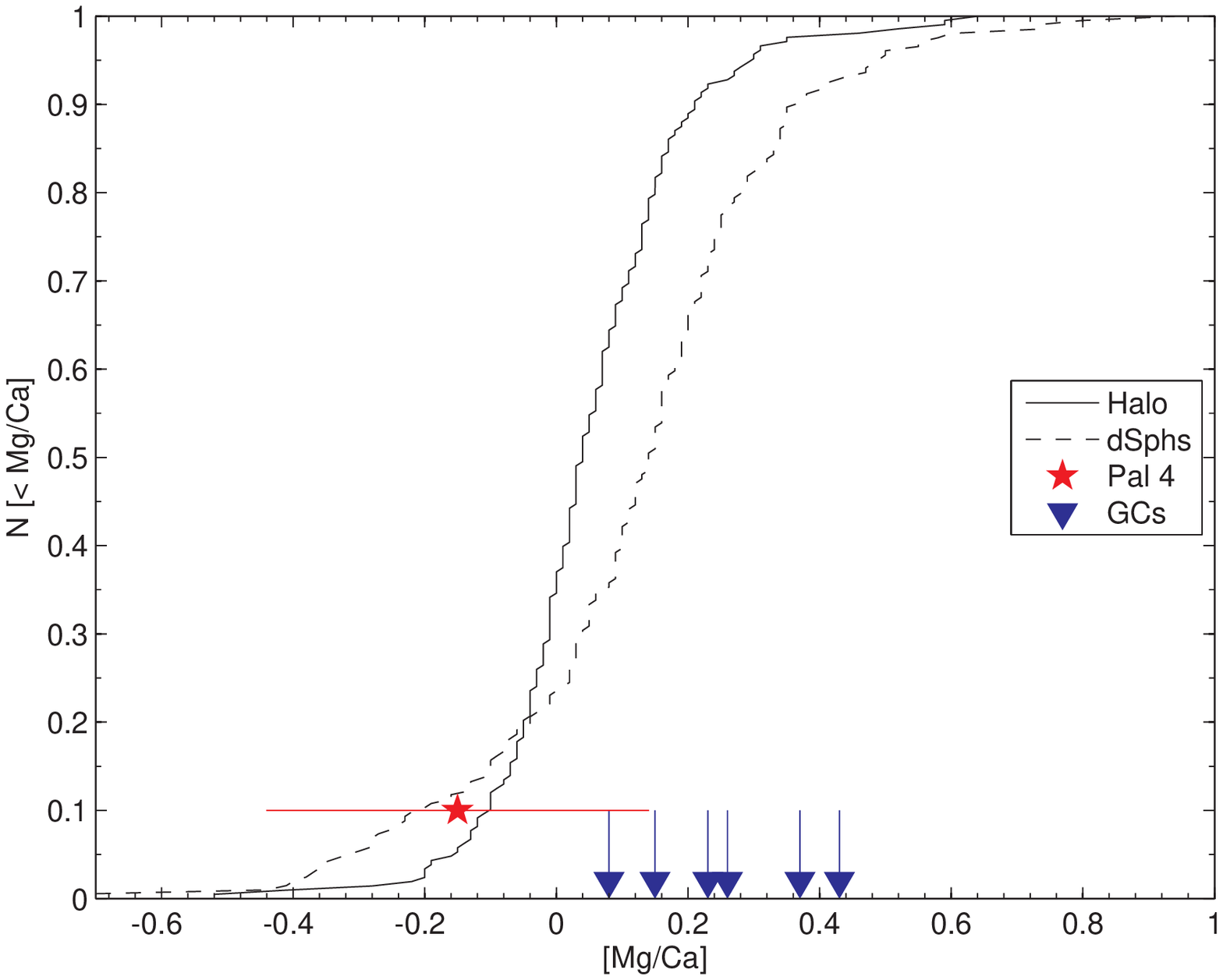}
\caption{Histograms (top panel) and cumulative distribution (bottom panel) of the [Mg/Ca] abundance ratio in Galactic halo stars (shaded histogram/black solid line) and dSph galaxies (open histogram/dashed line). Also indicated are the measurements for Pal~4 and  the Galactic GCs listed in Table~5 (see \S~5.4). The error bar on the Pal~4 data point is the squared sum of the total Mg and Ca/Fe errors. }
\end{figure}
\subsection{Iron peak elements}
Our measured [Sc/Fe], [Mn/Fe] and[ Ni/Fe] ratios are shown in Fig.~7.
\begin{figure}[htb]
\centering
\includegraphics[width=1\hsize]{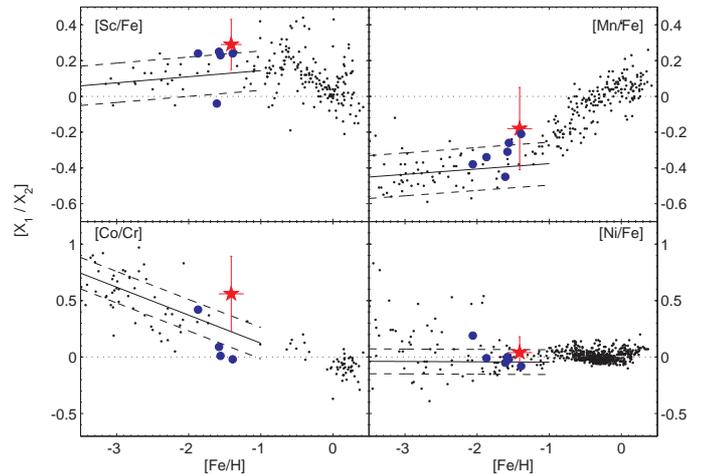}
\caption{Same as Fig.~5, but for [Sc/Fe], [Mn/Fe], [Ni /Fe] and [Co/Cr]. Black lines denote the regression lines and 1$\sigma$ scatter adopted from Cayrel et al. (2004), extrapolated to [Fe/H] = $-$1 dex.}
\end{figure}
Owing to the relatively large number of available Ni absorption lines, [Ni/Fe] is the best determined of these ratios, and, at 0.04 dex, has a value that is fully compatible 
with the Solar value (that is found over a broad range of iron abundances). This is not unexpected, since the iron-peak elements strictly trace the iron 
production in the long-lived SNe~Ia.  Cr is underabundant with respect to Fe, but fully compatible with Galactic halo stars, while [Co/Fe] is slightly higher 
than halo stars at the same metallicity. In Fig.~7, we choose to plot [Co/Cr] as this abundance ratio has proven to be relatively insensitive to systematic 
effects in the stellar parameters (e.g., McWilliam et al. 1995). The high Co abundance in Pal~4, coupled with a relatively low Cr abundance, leads to the marginally higher [Co/Cr] ratio indicated in this figure. Given its large uncertainty, and because we cannot rule out the possibility that this ratio has been affected by non-LTE effects, we will refrain from drawing any conclusions about the contributions  from massive stars yields to these elements' production in Pal~4 (cf. McWilliam et al. 1995; Koch et al. 2008a). 

Likewise, the [Mn/Fe] ratio in Pal~4, at $-$0.18 dex, is marginally  higher than the value of $\approx -0.4$~dex found for halo stars in the same [Fe/H] interval  
(for which we supplemented the plot with data from  Gratton 1989; Feltzing \& Gustafsson 1998;  Prochaska et al. 2000;  Nissen et al. 2000;  Johnson 2002, and Cayrel et al. 2004; see also McWilliam et al. 2003).  However, an intercomparison of Mn data usually suffers from zero point uncertainties  (e.g., McWilliam et al. 2003) in that abundances derived from the $\sim$4030 \AA\ triplet lines are systematically lower by 0.3--0.4 dex on average relative to the redder, high-excitation lines we employed in this study (e.g., Roederer et al. 2010). Thus Pal~4's elevated [Mn/Fe] does not appear unusual and we do not pursue this ratio any further. Finally, the [Cu/Fe] ratio (shown in Fig.~8 on top of the measurements in Galactic disk and halo stars by  Prochaska et al. 2000 and Mishenina et al. 2002) 
seem to agree well with the Galactic trend, suggestive of a common origin, although zero-point difficulties may also affect conclusions about the behavior of this element  (e.g., McWilliam \& Smecker-Hane 2005), 
as was the case for Mn. 
\subsection{Neutron capture elements}
We show in Fig.~8 the [Y/Fe] and [Ba/Fe] ratios as representatives of the heavy elements. 
\begin{figure}[htb]
\centering
\includegraphics[width=1\hsize]{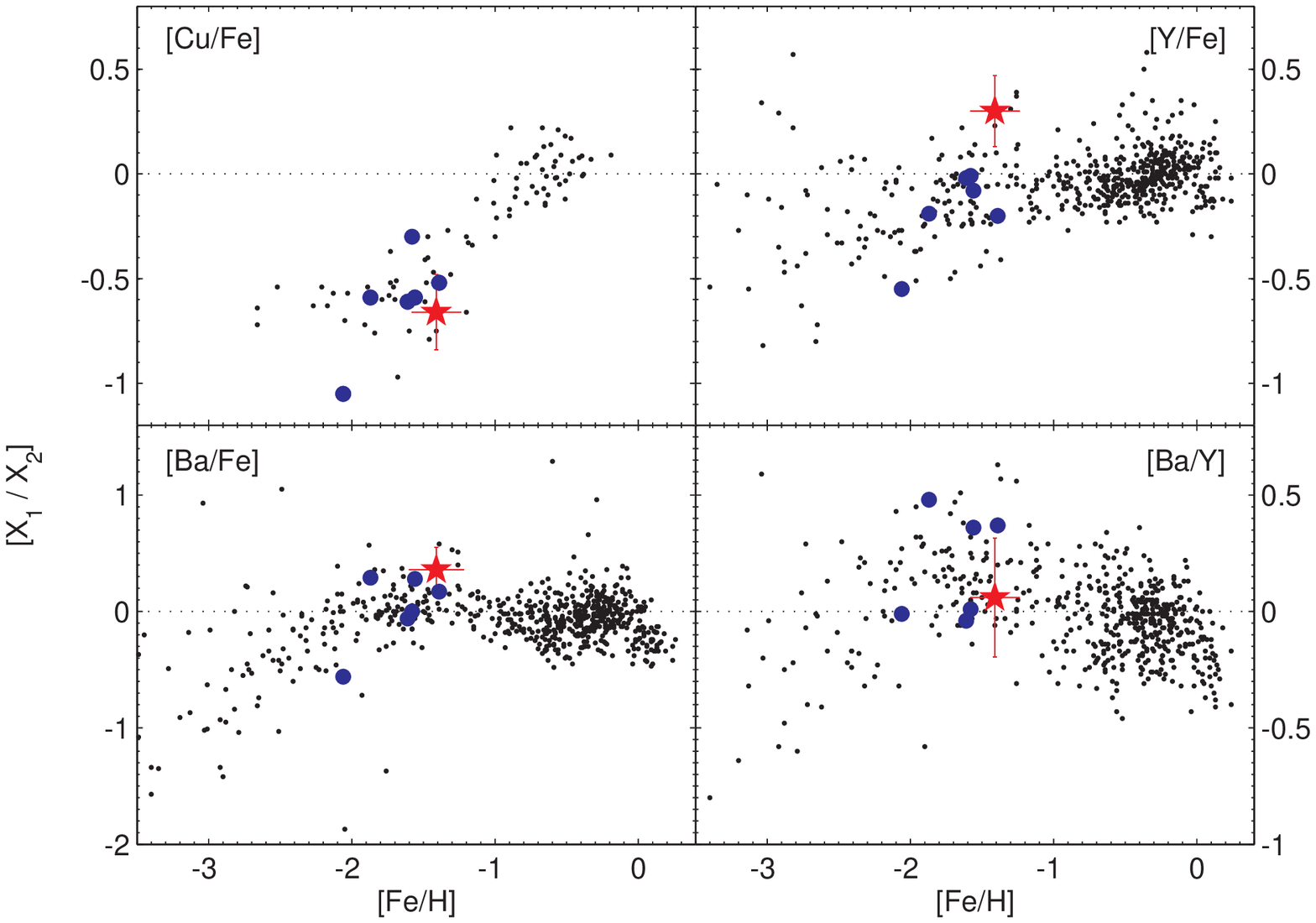}
\caption{Same as Fig.~7, but for [Cu/Fe], [Y/Fe], [Ba /Fe] and the $s$-process abundance ratio [Ba/Y].}
\end{figure}

All the elements with Z$>$38 are markedly enhanced relative to Fe. 
Unfortunately, our spectra lack information about the $r$-process element Eu, which prohibits any conclusions about the relative contributions of the AGB stars that 
produce the $s$-process elements to the early $r$-process production (most likely in massive SNe~II). 
On the other hand, the [Ba/Y] of $\sim$0.06 is fully compatible with the values found in Galactic halo stars, while it is strongly enhanced in the majority of the dSph stars studied to date owing to the importance of metal-poor AGB yields in the slow chemical evolution in these low-mass systems (e.g., Shetrone et al. 2003; Lanfranchi et al. 2008). 

Fig.~9 shows the heavy element abundances for Pal~4 together with the solar $r$-, $s$- and total scaled solar abundances from Burris et al. (2000). We have normalized 
the curves to the same Ba abundance. 
Unlike Pal~3, which was found to exhibit interesting evidence for a pure $r$-process origin, Pal~4's abundance data fall between  
 the $r$-process curve and the solar $r$+$s$-mix. However, the majority of these elements provide upper limits at most, so we refrain from a deeper discussion of the heavy element nucleosynthesis in this GC. 
\begin{figure}
\centering
\includegraphics[width=1.1\hsize]{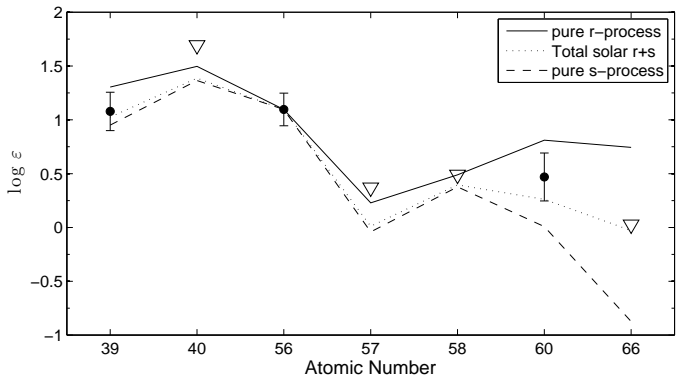}
\caption{Neutron capture elements in Pal~4, normalized to Ba. Black lines display the solar $r$- and $s$-process contributions from Burris et al. (2000). 
Triangles indicate upper limits.}
\end{figure}
\section{Comparison to Galactic Halo Tracers}
\subsection{Halo Globular Clusters}
Figure~10 and Table~5 compare our abundances for Pal~4 to those of a subsample of  Galactic GCs using data taken from the literature. 
Here we do not aim for a comprehensive comparison with the entire MWGC population (e.g., Pritzl et al. 2005; Geisler et al. 2007). Rather, we 
wish to simply compare Pal~4 to a few clusters that have been selected as broadly representative of the inner and outer halo cluster systems. 
Specifically, we use data for M3 and M13 from Cohen \& Mel\'endez (2005b),  which are archetypical {\em inner} halo GCs at R$_{\rm 
GC}\sim$9, 12 kpc (Z$_{\rm max}\sim$ 9,15 kpc) with metallicities similar to those of Pal~4. We also include 
NGC6752 in this comparison  as one of the nearest, inner halo clusters at a comparable metallicity (Yong et al. 2008). Finally, 
we include the {\em outer} halo clusters NGC7492 (Cohen \& Mel\'endez 2005a) and Pal~3 (Paper~I) as rare examples of remote
clusters with published abundances, as well as NGC5694, a GC that has been claimed to show abundance patterns more typical 
of dSph stars than GCs (Lee et al. 2006). 
\begin{figure}[htb]
\centering
\includegraphics[width=1\hsize]{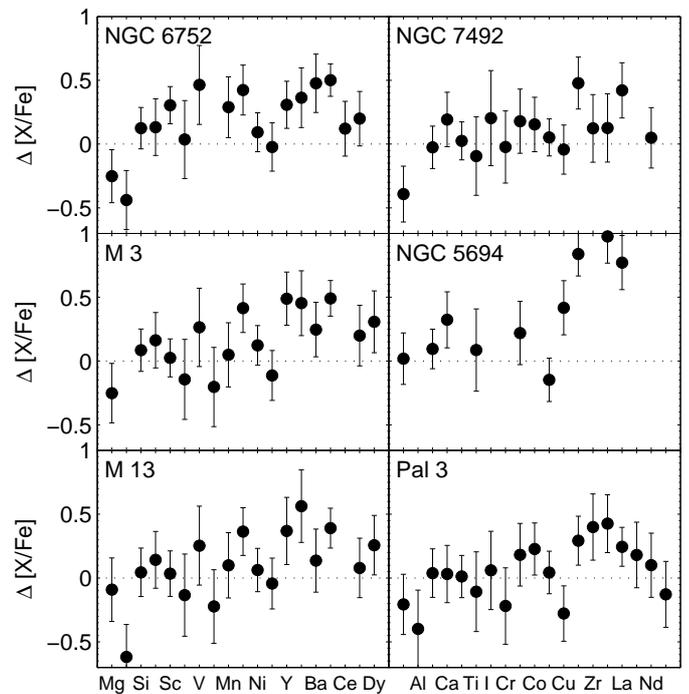}
\caption{Abundance differences for Pal~4, in the sense [X/Fe]$_{\rm Pal 4} - $ [X/Fe]$_{\rm GC}$. The six GCs are results assembled from the literature and corrected for different Solar abundance scales. Details for the GCs used in this comparison are given in Table~5.  Error bars include the 1$\sigma$-spreads from both this work and from the reference GC abundance ratios. For clarity, alternating labels are shown.}
\end{figure}

Table~5 shows the mean deviation $\langle\Delta$[X/Fe]$\rangle$ of Pal 4's abundance ratios from the literature values for the GCs chosen for reference. The fourth column lists the number of chemical elements, $N$, that the different studies have in common. 
Since Figs.~8 and 9 indicate that Pal~4 exhibits relatively high heavy element ratios with regard to the reference sample, we also computed the statistics for elements with Z$<$39 (Y) only. 
\begin{table*}
\caption{Pal~4 Abundances Relative to Comparison Clusters.}
\centering          
\renewcommand{\footnoterule}{} 
\begin{tabular}{rccrccccc}
\hline
\hline
Name & [Fe/H] & R$_{\rm GC}$ [kpc] & $N$ & $\langle\Delta$[X/Fe]$\rangle_{\rm all}$   & $\langle\Delta$[X/Fe]$\rangle_{\rm Z<39}$ & 
${\langle}{\Delta}$[X/Fe]/${\sigma}{\rangle}_{\rm all}$   & $\langle\Delta$[X/Fe]/${\sigma}{\rangle}_{\rm Z<39}$ & Reference$^a$\\ 
(1) & (2) & (3) & (4) & (5) & (6) & (7) & (8) & (9) \\
\hline
NGC 6752 & $-$1.61 &  5.2 & 17 & 0.18$\pm$0.06 &    \phs0.10$\pm$0.08 & 1.0 &    \phs0.5 & (1), (2) \\
M 13     & $-$1.56 &  8.7 & 17 & 0.15$\pm$0.06 &     \phs0.04$\pm$0.06 & 0.8 &    \phs0.2 & (1), (3) \\
M 3      & $-$1.39 &  12.3 & 18 & 0.09$\pm$0.06 & $-$0.01$\pm$0.08 & 0.5 &    \phs0.0 & (1), (3) \\
NGC 7492 & $-$1.87 & 24.9 & 16 & 0.09$\pm$0.05 &     \phs0.02$\pm$0.05 & 0.4 &    \phs0.1 & (1), (4) \\
NGC 5694 & $-$2.06 & 29.1 & 10 & 0.36$\pm$0.12 &     \phs0.14$\pm$0.06 & 1.8 &    \phs0.6 & (1), (5) \\
Pal 3    & $-$1.58 & 95.9 & 19 & 0.05$\pm$0.05 & $-$0.05$\pm$0.06 & 0.3 & $-$0.1 & (1), (6) \\
\hline
 \hline                  
\end{tabular}
\begin{list}{}{}
\item[$^a$] (1) web-version (2003) of Harris (1996); (2) Yong et al. (2005); (3) Cohen \& Mel\'endez (2005b); (4) Cohen \& Mel\'endez (2005a); (5) Lee et al. (2006);  (6) Paper~I.
\end{list}
\end{table*}
This comparison suggests that Pal~4 is, on average, enhanced with respect to each GC considered here if we account for all elements. On the other hand, the differences are statistically insignificant if we restrict the comparison to elements lighter than Y.  

Two of the comparison GCs in Table~5 show interesting discrepancies.  
The first, NGC6752, is the innermost object in the comparison sample and only slightly more metal-poor than Pal 4. Although its abundance patterns are similar to  the comparison GCs and field stars at this metallicity, Yong et al. (2005, and references therein)  found significant variations in the light {\em and} heavy elements, which supports the view that AGB stars alone cannot have carried the enrichment in the proto-cluster medium, although they likely played a significant role as indicated by the observed [Ba/Eu] ratios. While the observed differences for Z $\ga$ 39 would seem to suggest that the respective processes differed between the inner (NGC6752-like) and outer (Pal~4-like) halo, it is clear that more measurements --- particularly Eu abundances --- are needed.      

The second noteworthy example, NGC5694,  
exhibits heavy element abundance ratios that are incompatible with those of Pal~4. On the other hand, while its [Ca/Fe] is also significantly lower, we find an identical [Mg/Fe] ratio in Pal~4 (which, in turn, reflects in the different [Mg/Ca] ratios; see  Fig.~6). The low values of the $\alpha$-element ratios with respect to the Galactic halo have prompted Lee et al. (2006) to conclude that this GC is likely of an extragalactic origin. We shall return to this issue below.
\subsection{Halo Field Stars}
Is it safe to conclude that Pal~4 is typical of the Galactic halo population? In view of the relatively large errors that arise from the integrated nature of our analysis, we follow Norris et al. (2010) in first considering the {\em mean} halo abundance distribution. 
To this end, we computed the mean and dispersion for the Galactic halo and disk stars (shown as small black dots in Figs.~5, 7 and 8)  
as a function of [Fe/H] and fitted these relations with straight lines. Although this is an obvious oversimplification,
the halo data is adequately represented with these linear relations. The 
resulting range in the Galactic abundance ratios is shown by the black lines in Fig.~5. We emphasize that no efforts have been taken to homogenize the various data with respect to different approaches used in the analyses (i.e., regarding log $gf$ values and atmospheres), although we did correct for differences in the adopted Solar abundance scales when necessary.  
Note that Cayrel et al. (2004) also provide regression lines for [X/Fe] versus [Fe/H] based on their 35 metal-poor halo stars, but those stars have [Fe/H]$<-2.1$ dex and an extrapolation to metallicities of Pal~4 yields  slopes that are too high to describe the $\alpha$-element abundances shown here.

Pal~4 falls squarely on the regression lines for all $\alpha$-elements, except for Ca, although it is still consistent within the errors even in this case. Indeed, Pal~4 is generally in good agreement with the GCs shown in this comparison, with the exceptions noted above. In this picture, the proto-GC cloud from which Pal~4 formed was considerably enriched by the short-lived SNe~II that produced the $\alpha$-elements on rapid time scales --- a generic characteristic of the halo field stars and its genuine GC system. 

 For the even- and odd-Z iron-peak element ratios shown in Fig.~7, an extrapolation of the regression lines of Cayrel et al. (2004) provide good representations of the overall  halo trends  up to the metallicity regime around Pal~4 and higher. 
As argued above, the [Ni/Fe] ratio is well determined in Pal~4 and is fully compatible with the Solar value that is observed in halo field and GC stars,  
 bolstering the ubiquity of iron-peak nucleosynthesis in the SNe~Ia at [Fe/H] above $\sim-2$ dex. 
The [Sc/Fe] ratio in Pal~4 falls towards the upper limit of the halo distribution, which holds for all the reference GCs in our sample, except NGC~6752. Likewise, the slight Mn enhancement is not atypical and agrees well with, for instance, M3. This may indicate that the metallicity dependence of the SNe~II yields was less pronounced in the Pal~4 proto-GC cloud (cf. McWilliam et al. 2003). 
Finally, we note that the [Co/Cr] ratio is significantly larger that those observed for the three GCs in this metallicity range. These GCs show roughly solar values, as would be expected as both elements fall close together on the iron peak. McWilliam et al. (1995) first detected a strong rise of this ratio in metal-poor halo stars below 
$\sim-2.4$ dex. In fact, the observed [Co/Cr] of 0.55$\pm$0.33 dex is reminiscent of NGC~7492, albeit at a 
metallicity that is higher by roughly 0.5~dex. 

In the case of Cu, and the n-capture elements Y and Ba, the scatter in the halo abundance ratios is more difficult to evaluate due to a much sparser sampling of those elements and a notably increased (and real) abundance scatter among the metal-poor stars below $\sim -2$~dex. 
We therefore restrict the following brief discussion of Fig.~8 to the scatter plots without quantifying any linear trends. 

While the [Y/Fe] ratio lies above the bulk of the halo data, and is also higher than our comparison clusters by more than 0.3~dex, Ba seems only mildy enhanced with respect to these populations. Overall, the $s$-process ratio [Ba/Y] is in full agreement with the halo fields stars within the scatter. 
However, Pritzl et al. (2005) have shown that, in comparison with (thick) disk GCs, the halo clusters tend to be offset more towards higher [Ba/Y] ratios, and so are the dSphs. The latter is usually interpreted in terms of the low star formation efficiencies of the dSph galaxies, which leaves room for  a much stronger contribution from 
metal-poor AGB stars that are the main  sites of the $s$-process (e.g., Busso et al. 2001; Lanfranchi et al. 2008). 
The three GCs with the very high [Ba/Y] ratios in Fig.~10 are M3, M13 and NGC7492 and therefore representatives of the inner {\em and} outer halo. Following this line of reasoning, the 
slow star forming rates and metallicity dependent AGB-yields that cause enhancements in this ratio appear to be unrelated to location within the halo. 
In Paper~I we found that Pal~3's heavy elements are largely governed by $r$-process nucleosynthesis. From the sparse data for Z$>$38 in NGC~5496, it cannot be excluded that this cluster also follows this trend, so that the above arguments regarding bimodal $s$-process ratios may not apply to these remote halo clusters.  In any case, we emphasize that detailed $r$- and $s$-process abundance measurements for individual stars are vital for resolving these questions.  

\subsection{Comparison with other Substructures in the Outer halo}

In this section, we compare our abundances for Pal~3 (Paper~I) and Pal~4 to published values for other ``substructures" or ``overdensities" in the outer halo of the Milky Way, regardless of their morphological classification. Our comparison therefore focuses on a sample of 13 halo GCs, seven dSph galaxies (Sagittarius, Fornax, Draco, Sextans, Carina, Ursa Minor and Leo II) with abundance data from 
Shetrone et al. (2001; 2003; 2009); Sadakane et al. (2004); Monaco et al. (2005); Letarte (2007); Koch et al. (2008b); Cohen \& Huang (2009); Aoki et al. (2009); 
 and the five so-called ``ultra-faint" dSph galaxies (hereafter UF-dSphs; Hercules, Coma Berenices, Ursa Major~II, Bootes~I and Leo~IV) with published abundance information (Koch et al. 2008a; Frebel et al. 2010; Feltzing et al. 2009; Simon et al. 2010). 

All GCs shown here were selected to have Galactocentric distances $R_{GC} \gtrsim 8$~kpc; including Pal~3 and Pal~4 gives us a total of five GCs beyond $R_{GC} = 25$~kpc, and three of these GCs (Pal~3, Pal~4 and NGC2419) are at $R_{GC} \ge 90$~kpc. Note that only two other GCs in the catalog of Harris (1996) lie at or beyond this distance (Eridanus and AM~1). Thus, while the available abundance measurements are certainly still sparse (i.e., being based on just a single RGB star in NGC2419, four RGB stars in Pal~3, and co-added spectra for 19 RGB stars in Pal~4; Shetrone et~al. 2001; Paper~I), it is now possible to have first glimpse into the abundance patterns of the most remote Galactic GCs, and their relationship, if any, to the dSph and UF-dSph galaxies residing in the outer halo.
Because the number of element abundance measurements is generally limited (and differs amongst the various studies), we restrict our comparison to [Fe/H] and [$\alpha$/Fe], where we take [$\alpha$/Fe] $\equiv$ ([Mg/Fe] + [Ca/Fe] + [Ti/Fe])/3. 
\begin{figure*}[htb]
\centering
\includegraphics[width=1.07\hsize]{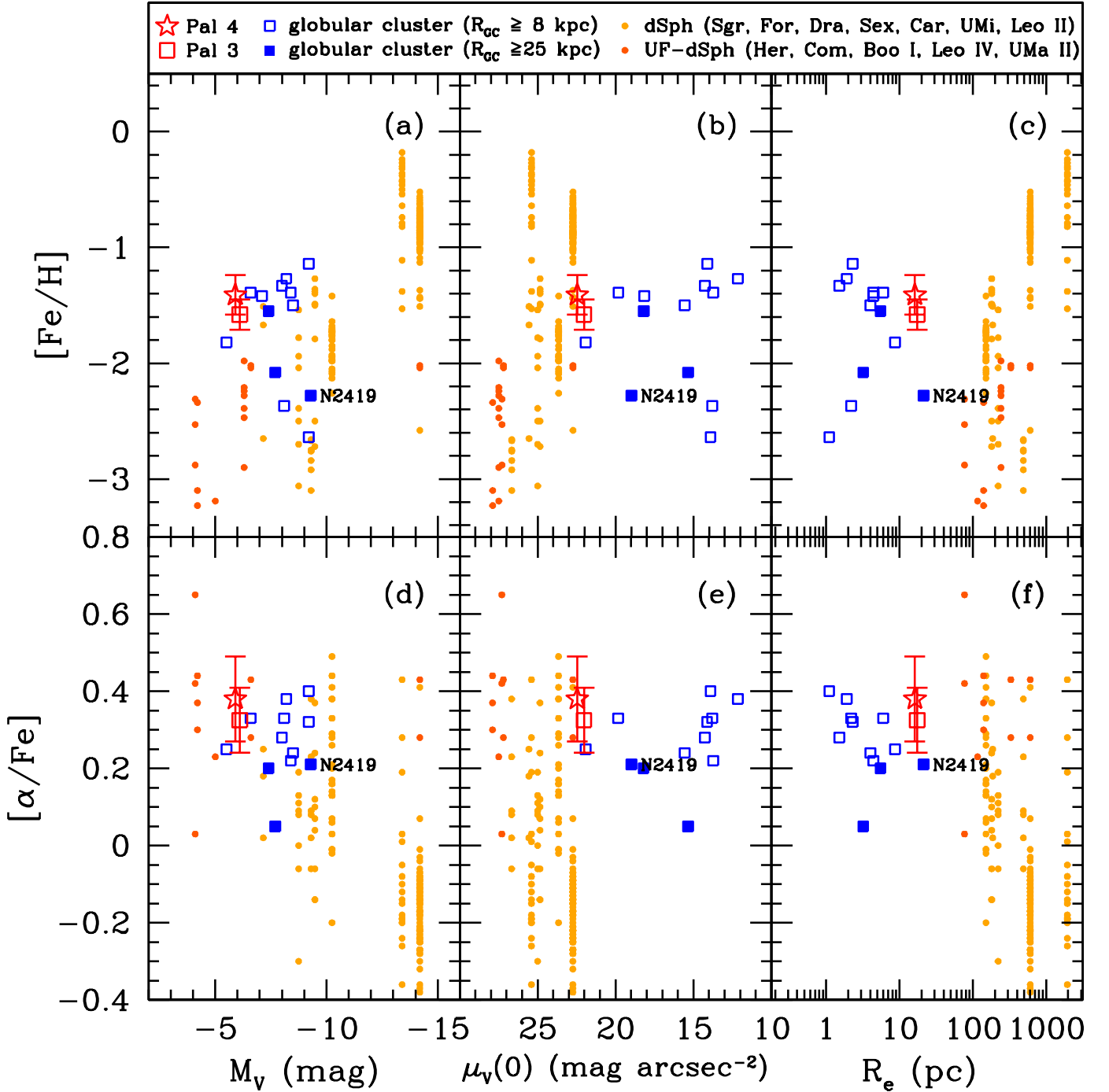}
\caption{Dependence of [Fe/H] and [$\alpha$/Fe] on structural parameters for various types overdensities in the Galactic halo: globular clusters (open and filled blue squares), dSph galaxies (orange circles) and ``ultra-faint" dSph galaxies (brown circles). The large red symbols show our results for Pal 3 (Paper~I) and Pal~4 (this paper). The luminous globular cluster NGC2419, which lies at a Galactocentric distance comparable to Pal~3 and Pal~4, is labeled in each panel. The structural parameters shown in this figure are absolute $V$-band magnitude {\it (panels a,d)}, central $V$-band surface brightness {\it (panels b,e)} and effective radius {\it (panels c,f)}.}
\end{figure*}

In Fig.~11, we show the behaviour of [Fe/H] and [$\alpha$/Fe] for stars belonging to halo GCs (blue squares), the more luminous dSph galaxies (orange circles) and UF-dSph galaxies (brown circles). Note that we plot GCs in the range $8 \ge R_{GC} \ge 25$~kpc as open blue squares, while GCs in the range $R_{GC} \ge 25$~kpc are shown as filled blue squares. Abundances are plotted against total $V$-band magnitude, $M_V$, central $V$-band surface brightness, $\mu_V(0)$, and effective (or half-light) radius, $R_e$ (Harris 1996; Irwin \& Hatzidimitriou 1995; Mateo 1998; McLaughlin \& van der Marel 2005; Martin et al. 2008). Pal~3 and 4 are highlighted as the large red square and star, respectively, while the third GC at $R_{GC} \gtrsim 90$~kpc, NGC~2419, is labelled in each panel. 

There are several interesting conclusions  to be drawn from this figure. First, Pal~3 and 4 appear as near ``twins" in this comparison, having similar Galactocentric distances, structural parameters (notably large radii), $V$-band luminosities, metallicities and $\alpha$-element enhancements. NGC2419, although much more luminous than either Pal~3 or Pal~4, appears similar in terms of its $\alpha$-enhancement. For these three GCs, which lie in the range $91 \lesssim R_{GC} \lesssim 112$~kpc, we find a mean of
$$\langle{\rm {[}}\alpha/{\rm Fe]}\rangle  = +0.31\pm0.09~{\rm dex}.$$
Adding NGC5694 and NGC7006, we find
$$\langle{\rm {[}}\alpha/{\rm Fe]}\rangle  = +0.24\pm0.13~{\rm dex}$$
for the five GCs with $R_{GC} \gtrsim 25$~kpc.
Thus, on the whole, Pal~3 and Pal~4 seem to have levels of $\alpha$ enhancement that are similar to most other halo GCs and nearby halo field stars, but slightly higher than dSphs at comparable metallicities (e.g., Shetrone et~al. 2001, 2003; Venn et~al. 2004; Koch 2009).
It is important to bear in mind, however, that stars in individual dSph galaxies show significant scatter, and it is certainly true that some dSph stars fall close to the region in the [Fe/H]--[$\alpha$/Fe] diagram occupied by these remote GCs: i.e., 10/157 $\approx$ 6\% of the dSph stars plotted in Fig.~11 fall within the 2$\sigma$ uncertainties for Pal~3 and Pal~4. 

In absolute terms, the {\em mean} [$\alpha$/Fe] for the most remote GCs is indistinguishable from that found in the UF-dSph galaxies shown in Fig.~11, which have $+0.36\pm0.17$~dex and a full range of $+0.03$ to $+0.65$~dex (based on measurements for nine stars in Her, UMa~II, Com, and Leo~IV). Note that Pal~3 and 4 are atypical of Galactic GCs in terms of their structural parameters,  being unusually extended ($R_e \gtrsim 15$~pc, or roughly fives times as large as ``typical" GCs; Jord\'an et~al. 2005) and having low surface brightness (with $\mu_V(0) \gtrsim 22-22.5$~mag~arcsec$^{-2}$). Thus, at least superficially, these remote GCs may have more in common with some UF-dSph galaxies than their apparent counterparts in the inner halo.  

There are, at present, two characteristics of the UF-dSph population that suggest they are indeed low-luminosity galaxies rather than faint, extended GCs (e.g., Larsen \& Brodie 2002; Mackey et~al. 2006; Peng et~al. 2006). The first such characteristic is their very large mass-to-light ratios, which point to the presence of significant dark matter halos (Simon \& Geha 2007; Strigari et~al. 2008). Secondly, the UF-dSphs seem to have abundances that fall along the extrapolation of the dwarf galaxy metallicity-luminosity relation, with significant intrinsic dispersions in metallicity (Kirby et~al. 2008). Using these criteria, what can we conclude about the origin of the most remote halo GCs?

Unfortunately, it is difficult to draw firm conclusions on possible metallicity spreads in these systems since there are measurements for just a single RGB star in NGC2419 (Shetrone et~al. 2001), and our analysis of co-added spectra in Pal~4 presupposes that there is no abundance spread (see \S1; \S5.2.). In the case of Pal~3, where high-quality MIKE spectra are available for four RGB stars, we can confidently rule out an abundance spread larger than $\sim$~0.1~dex (Paper~I). Regarding the dark matter content of these systems, Baumgardt et~al. (2009) have recently carried out a dynamical analysis of NGC2419, finding $M/L_V = 2.05\pm0.50$ in solar units. This value is typical of GCs (McLaughlin \& van der Marel 2005) and {\it much} smaller than the extreme values reported for UF-dSphs (e.g., Simon \& Geha 2007; Strigari et~al. 2008). Detailed dynamical modeling of Pal~3 and 4 will be the subject of a future paper in this series, but it is clear that the extreme $M/L_V$ values for UF-dSph galaxies can be ruled out at a very high confidence (i.e., for a system like Pal~4, with $L_V \sim 2.1\times10^4~L_{V,{\odot}}$, known UF-dSphs have mass-to-light ratios of $\approx$ $10^3$ to 10$^4$; Strigari et~al. 2008; Geha et~al. 2009).

In short, the available evidence suggests that Pal~4 (and Pal~3) formed in a manner resembling that of typical halo GCs, although it is clear that additional abundance measurements for stars in these and other remote GCs is needed urgently. Indeed, each contains many RGB stars that are well within the reach of high-resolution spectrographs on 8m-class telescopes. Such observations would allow a direct measurement of the intrinsic abundance spread within these systems --- an important clue to their origin and relationship to other halo substructures such as dSph and UF-dSph galaxies.

\section{Summary}
Motivated by the good agreement between the abundance ratios measured from high-S/N spectra of individual stars in Pal~3 and those found using co-added, low-S/N spectra (Paper~I),
we have used the same technique to measure chemical abundance ratios in the remote halo GC Pal~4. Although systematic uncertainties and the low S/N ratios complicate such studies, an accuracy of 0.2 dex is possible for most abundance ratios, sufficient to place such faint and remote systems into a context with both the inner and outer halo GCs, as well as dSph and UF-dSph galaxies. In the future, this technique may enable the global abundance patterns to be characterized in additional remote systems, allowing a first reconnaissance of the chemical enrichment histories of remote Galactic satellites. 

Perhaps the most striking finding in Pal~4 is the subsolar [Mg/Ca] ratio, which is not observed in the sample of reference GCs that span a broad range of Galactocentric distances. Despite an overlap of our observed ratio with the halo field population, its low value may rather 
resemble the low-[Mg/Ca] tail of the distribution for dSph stars. In contrast, we see tentative evidence for a solar [Ba/Y] ratio, which militates against a slow chemical evolution and accompanying AGB enrichment as suggested by enhanced [Ba/Y] values in about two thirds of the dSph stars studied to date. 
Overall, most of the element ratios determined in this study overlap with the corresponding measurements for halo field stars, although a few ratios seem to fall above the halo star trends (see \S5). This favors a scenario in which the material from which both Pal~4 and the Galactic halo formed underwent rather similar enrichment processes.  

In their analysis of the CMD of Pal~4, Stetson et al. (1999) state that the cluster is younger than the inner halo GC M5 by about 1.5 Gyr (at [Fe/H]=$-$1.33 dex; Ivans et al. 2003; Koch \& McWilliam 2010) {\em if} they ``all have the same composition -- and [...] this means both [Fe/H] and [$\alpha$/Fe]''. Our work 
has shown that Pal~4 is enhanced by +0.38$\pm$0.11~dex in the $\alpha$-elements, which is  consistent with the value of 0.3 dex   assumed in the above CMD modeling. On the other hand, the CMD analysis suggested an [Fe/H] of $-$1.28 dex, which is slightly more metal rich than what we found in the present spectroscopic study: $\langle$[Fe/H]$\rangle$ = $-$1.41 dex. As noted in Vandenberg (2000), ``an increase in [Fe/H] or [$\alpha$/Fe] would result in slightly younger [...] ages'' for Pal~4 (as determined via the magnitude offset between the horizontal branch and main-sequence turnoff). This would imply that Pal~4 is slightly older than found in Stetson et~al (1999) and hence more similar in age to the older halo population. This, however, is in contradiction to the younger age suggested by its peculiar (i.e., red) horizontal branch morphology, unless further parameters, such as red giant mass loss, are invoked  (Catelan 2000). 

Based on the evidence at hand, Pal~4 seems to have an abundance pattern that is typical of other remote GCs in the outer halo. An open question, given the nature of our analysis which relies on co-adding individual RGB star spectra, is whether Pal~4 is monometallic or, like dSph and UF-dSph galaxies, shows an internal spread in metallicity. We argued in Sect.~5.2. judging from our limited quality spectra, however, that it is unlikely that this object exhibits any significant intrinsic iron scatter.  
It is clear that high-quality abundance ratio measurements for individual stars in Pal~4 and other remote substructures are urgently needed to understand the relationship, if any, between remote GCs and other substructures in the outer halo.

 \begin{acknowledgements}
We thank I.U. Roederer for discussions and an anonymous referee for a very helpful report. 
AK acknowledges support by an STFC postdoctoral fellowship. This work was based on observations obtained at the W. M. Keck Observatory, which is operated jointly by the California Institute of Technology and the University of California. We are grateful to the W. M. Keck Foundation for their vision and generosity. We recognize the great importance of Mauna Kea to both the native Hawaiian and astronomical communities, and we are grateful for the opportunity to observe from this special place.
 \end{acknowledgements}

\end{document}